\documentclass[useAMS,usenatbib]{mn2e}

\usepackage{psfig}
\usepackage[dvips]{graphicx}
\usepackage{lscape}
                                 
\def\lsim{\mathrel{\hbox{\rlap{\hbox{\lower4pt\hbox{$\sim$}}}\hbox{$<$}}}}
\def\gsim{\mathrel{\hbox{\rlap{\hbox{\lower4pt\hbox{$\sim$}}}\hbox{$>$}}}}

\title[Fully cosmological virtual massive galaxies at $z=0$]{Fully cosmological virtual massive galaxies at 
$z=0$: kinematical, morphological, and stellar population characterisation}

\author[Javier Navarro-Gonz\'alez et al.]{Javier Navarro-Gonz\'alez$^{1}$,
 Elena Ricciardelli$^{1}$\thanks{e-mail: elena.ricciardelli@uv.es} , Vicent Quilis$^{1}$, Alexandre Vazdekis$^{2,3}$\\
$^{1}$Departament d'Astronomia i Astrof\'{\i}sica, Universitat de Val\`encia,
c/ Dr. Moliner 50, E-46100 - Burjassot, Val\`encia, Spain\\
$^{2}$Instituto de Astrof\'{\i}sica de Canarias, 
c/ V\'{\i}a L\'actea s/n, E38205 - La Laguna, Tenerife, Spain\\
$^3$Departamento de Astrof?õsica, Universidad de La Laguna, E-38205,
Tenerife, Spain}

\begin{document}
\date{Accepted ...  Received ...; in original form ...
}
\pagerange{\pageref{firstpage}--\pageref{lastpage}} \pubyear{...}
\maketitle
\label{firstpage}

\begin{abstract}
We present the results of a numerical adaptive mesh refinement hydrodynamical and N-body simulation in a $\Lambda CDM$ 
cosmology. We focus on the analysis of the main properties of massive 
galaxies ($M_* > 10^{11}\,M_{\odot}$) at $z=0$. 
For all the massive virtual galaxies 
we carry out a careful study of their one dimensional density, 
luminosity, velocity dispersion, and stellar population profiles. 
In order to best compare with observational data, the method to estimate the velocity dispersion
is calibrated by using an approach similar to that performed in the
observations, based on the stellar populations of the simulated
galaxies. With these ingredients, 
we discuss the different properties of massive galaxies in our sample 
according to their morphological types, accretion histories and dynamical 
properties.  We find that the galaxy merging history is the leading actor in
shaping the massive galaxies that we see nowadays.
Indeed, galaxies having experienced a turbulent life are the most
massive in the sample and show the steepest metallicity gradients.
Beside the importance of merging, only a small fraction of the final
stellar mass has been formed ex-situ (10-50\%), while the majority of
the stars formed within the galaxy.
These accreted stars are significantly older and less metallic  than
the stars formed in-situ and tend to occupy the most external regions
of the galaxies.

\end{abstract}


\section{Introduction}

Understanding the formation and evolution of massive galaxies (i.e. galaxies with stellar mass $M_* >10^{11}
M_{\odot}$) represents one of the major challenge in the current
hierarchical model of galaxy formation. 
In the nearby Universe, the majority of massive galaxies have the
morphology of early-type galaxies, ETG, \citep{baldry04, renzini06}. 
The  bulk of their stars is old and metal-rich \citep{jorgensen99, Trager00,
  gallazzi05}, pushing them on the red sequence in
the colour-magnitude relation.
Though, several observational evidences also revealed that a small
amount of recent star formation is a common feature in massive
galaxies \citep{Trager00, Bressan06, Kav07, Sarzi08, Toj11}.

From the dynamical point of view, massive ETGs show the same dichotomy
as the general galaxy population. 
Integral-field kinematics of the  SAURON sample \citep{sauron}, 
has allowed to distinguish two families of ETGs:  
the slow rotators (SR), having little or no rotation, showing  
 misalignments between the photometric and the kinematic axes, and
 containing kinematically decoupled cores, 
and the fast rotators (FR), showing a disk-like rotation
\citep{cappellari07, emsellem07, 
  emsellem11}.
The fast rotators are the most populated family,  constituting the $
\sim 70\%$ of the ETGs brighter than $M_{K} \sim -24$.
Dissipative processes, like gas-rich mergers or gas accretion, are often invoked to explain the formation of
fast rotators \citep{bender92, bournaud05, bois12}. Dissipationless mergers are instead generally
assumed as the most likely
mechanism to produce slow rotators \citep{naab03, cox06}. The number
of major mergers during a galaxy life  can also play
a role in differentiating among fast and slow rotators \citep{khochfar11}.

In conjunction with their old stellar populations, even the assembly
process in massive galaxies appears to take place at moderately high
redshift, as indicated by the little evolution of the high-mass end of
the stellar mass function \citep{fontana06, pg08, marchesini09}. 
The existence of a population of massive, old, and passively
evolving galaxies, observed up to $z \sim 2$ \citep{cimatti04, glazebrook04,cimatti08,whitaker13}, 
has opened the question on how to
form such galaxies in a $\Lambda$CDM cosmology, where the structure
growth is expected to be hierarchical. 
In this context, the major merger scenario certainly plays a
significant role in shaping the present-day massive galaxies.  
Binary mergers of gas-rich disks has been shown to be a viable
mechanism for the  formation of spheroidal galaxies \citep{barnes92,
  hernquist92, cox06, naab06}. 
Although major mergers certainly occur and can explain the existence of
massive compact galaxies at high redshift \citep{ricciardelli10,
  bournaud11}, they are too rare and other mechanisms should be active to explain the
subsequent evolution.  
Minor mergers, being much more common \citep{khochfar09}, are likely to
provide this mechanism. They represent also the natural way 
to explain the strong size
evolution of ETGs observed between z=2 and the present time \citep{daddi05,
  trujillo07, cimatti08, cassasa11,damjanov11}, while, at the same time, lead to  a mild
decrease in the velocity dispersion, as observed \citep{cenarro09, naab09}.

The picture emerging from cosmological simulations of massive galaxies
\citep{naab09, oser10,  lackner12} naturally embeds the minor merger
scenario as a fundamental ingredient in  a two-phase formation process.
In a first phase, galaxies assemble their
mass mainly through dissipative processes and star formation takes
place in-situ. This in-situ star formation can be induced by cold flow 
accretion  \citep{Keres05, Ocvirk08, Keres09, dekel09}  or by 
gas-rich mergers  \citep{mihos96, cox08}.
 In a second phase, the
mass assembly occurs mainly by accretion of satellites.
In this phase, mergers occur with typical mass ratio of 1:5.
Simulations based on different codes agree in that the ex-situ
component is made by older and less metallic stars than the in-situ
population \citep{lackner12,johansson12}. Hydrodynamical simulations of disc galaxies also find important 
differences between the in-situ and the accreted components 
\citep{zolotov09, font11}.
However, the details on the two-phase galaxy formation are still
strongly model dependent and the mass fraction of accreted stars can
vary by more than 50\% among different models (see for instance
\citealt{lackner12}). Differences in the numerical techniques as well
as in the sub-grid physics can lead to such important discrepancies
 \citep{Dubois13}.

The purpose of this paper is to characterize a sample of simulated massive
galaxies from an adaptive mesh refinement (AMR) simulation. 
We study their zero-redshift properties, in terms of morphology,
kinematics and stellar population content and link them to the galaxy
merging history.
Following the framework of the two-phase galaxy
formation we separate the stars formed in-situ from those accreted and
study their properties.
We couple the outcome of the simulation with stellar population synthesis models in order to 
present our results in a form as closer as possible to observations, rather than using the raw data from 
the simulations.
The structure of the paper is as follows. In Section \ref{numerics} and \ref{shaping} we
describe the details of the simulations and of the tools used in the
post-processing analysis.  In Sections \ref{results} we present
one-dimensional profiles of the relevant quantities. In Section \ref{insitu} we describe the
properties of the stellar populations formed in-situ and
ex-situ. Finally, we draw our conclusions in Section
\ref{discussion}. In the appendix we discuss our method to measure the
galaxy velocity dispersion.

\section{Simulating the Virtual galaxies}\label{numerics}

\subsection{Numerical simulation}\label{simulations}

The  simulation  described  in  this  paper was  performed  with  the
cosmological  code  MASCLET \citep{quilis04}.   This  code couples  an
Eulerian  approach  based  on  {\it high-resolution  shock  capturing}
techniques  for describing  the  gaseous component,  with a  multigrid
particle mesh  N-body scheme for evolving  the collisionless component
(dark matter).  Gas and dark matter are coupled by the gravity solver.
Both  schemes  benefit of  using  an  adaptive  mesh refinement  (AMR)
strategy, which permits to gain spatial and temporal resolution.

The numerical  simulation was run  assuming a spatially  flat $\Lambda
CDM$  cosmology, with  the following  cosmological  parameters: matter
density    parameter,    $\Omega_m=0.25$;    cosmological    constant,
$\Omega_{\Lambda}=\Lambda/{3H_o^2}=0.75$;  baryon  density  parameter,
$\Omega_b=0.045$;  reduced Hubble  constant, $h=H_o/100  km\, s^{-1}\,
Mpc^{-1}=0.73$;  power  spectrum index,  $n_s=1$;  and power  spectrum
normalisation, $\sigma_8=0.8$.

The initial  conditions were  set up at  $z=50$, using a  CDM transfer
function from \citet{EiHu98},  for a cube of comoving  side length $44
\,  Mpc$.   The  computational  domain was  discretised  with  $128^3$
cubical cells.

Two  levels  of refinement (level $l=1,2$) for the  AMR scheme were set
up from the initial conditions by selecting regions satisfying certain
refining criteria,  when evolved  -- until present  time --  using the
Zeldovich  approximation.  The  dark matter  component in  the initial
refined  regions were  sampled with  dark matter  particles  eight 
and sixty four times, respectively,
lighter than  those used  in regions covered  only by the  coarse grid
(level $l=0$).   During the evolution, regions on  the different grids
are refined  based on  the local baryonic  and dark  matter densities.
The ratio  between the cell  sizes for a  given level ($l+1$)  and its
parent   level  ($l$)   is,   in  our   AMR  implementation,   $\Delta
x_{l+1}/\Delta  x_{l}=1/2$.  This  is a  compromise value  between the
gain in resolution and possible numerical instabilities.

The simulation presented  in this paper uses a  maximum of seven levels
($l=7$) of refinement, which gives  a peak physical spatial 
resolution of $\sim
2.7\,  kpc$ at $z=0$. For the 
dark  matter we  consider three  particles species,
which correspond to the particles on the coarse grid and the particles
within the two first  level of refinement at the  initial conditions.  The
best  mass resolution  is  $\sim 2\times  10^7\,  M_\odot$. This is 
equivalent to use $512^3$ particles in the whole box.

Our simulation includes cooling  and heating processes which take into
account inverse Compton and free-free  cooling, UV heating \citep{hama96}, and
atomic and molecular cooling for a primordial gas. In order to compute
the abundances  of each species, we  assume that the  gas is optically
thin  and in ionization  equilibrium, but  not in  thermal equilibrium
\citep{katz96,theuns98}.  The tabulated  cooling rates were taken from
\citet{sudo93} and they depend on the local metallicity.
The cooling curve was  truncated below temperatures of $10^4\,K$.  The
cooling and heating  were included in the energy  equation (see Eq.(3)
in \citet{quilis04}) as extra source terms.

\subsection{Star formation and chemical enrichment}\label{starformation}
 
The star  formation is  introduced in the  MASCLET code  following the
ideas  of \citet{yepes97}  and \citet{springe03}.   In  our particular 
implementation, we assume that cold  gas in a cell is transformed into
star  particles on  a  characteristic time  scale  $t_*$ according  to
$\dot{\rho_*}=-\dot{\rho}=(1-\beta)\,{\rho}/{t_*(\rho)}$  where $\rho$
and  $\rho_*$  are the  gas  and  star  densities, respectively.   The
parameter  $\beta$  stands for  the  mass  fraction  of massive  stars
($>8\,M_{\odot}$) that explode as  supernovae, and therefore return to
the  gas  component in  the  cells.   We  adopt $\beta=0.1$,  a  value
compatible with a Salpeter IMF.  For the characteristic star formation
time,        we        make        the        common        assumption
$t_*(\rho)=t^*_o(\rho/\rho_{_{th}})^{-1/2}$,       equivalent       to
$\dot\rho_* \sim \rho^{1.5}/t^*_o$   \citep{keni98}.   In   this   way,  we
introduce a dependence on the local  dynamical time of the gas and two
parameters, the density threshold for star formation ($\rho_{_{th}}$)
and  the corresponding  characteristic time  scale ($t^*_0$).   In our
simulation,       we        take       $t^*_0=2\,       Gyr$       and
$\rho_{_{th}}=10^{-25}\,g\,cm^{-3}$.  From  the energetic point
of view,  we consider that each  supernova dumps in  the original cell
$10^{51}\, erg$ of thermal energy. In a similar way, we assume that 
every time that a star forms, it returns to the environment a fraction of metals depending on its mass, 
$y=\frac{m_{_M}}{m_*}$, where $y$, $m_{_{M}}$, and $m_*$ are the yield, the mass of metals, and the 
star mass, respectively.
This metal density $(\rho_{_{M}})$ allows us to define a metallicity, $Z=\frac{\rho_{_{M}}}{\rho}$
which can be used to compute cooling rates for variable metallicities. The metallicity is advected 
through the computational box using a continuity equation similar to the continuity equation of 
the gas component.
 Our stellar formation approach does not take into account the feedback from stellar winds (AGB stars) 
nor the type Ia supernovae.

In the practical implementation,  we assume that star formation occurs
once every global  time step, $\Delta t_{l=0}$ and,  only in the cells
at the highest levels of refinement ($l=6,7$).  Those cells at these levels
of   refinement,  where  the   gas  temperature   drops  below   $T  <
2\times10^4\,    K$,    and   the    gas    density    is   $\rho    >
\rho_{_{th}}=10^{-25}\,  g\,  cm^{-3}$,  are suitable  to  form
stars.   In  these  cells,  collisionless  star  particles  with  mass
$m_*=\dot\rho_*\Delta  t_{l=0}\Delta x_l^3$ are  formed.  In  order to
avoid sudden changes in the  local gas density, an extra condition restricts
the    mass    of    the    star    particles    to    be    $m_*={\rm
min}(m_*,\frac{2}{3}m_{gas})$, where  $m_{gas}$ is the  total gas mass
in the considered cell.  The energy associated to the stellar feedback from supernovae is dumped
within the same cell where the stellar particle is created.  The adopted value for the yield is 
$y=0.02$.

\subsection{Stellar populations}

To convert physical quantities in observables, we employ the MIUSCAT stellar
population models  \citep{vazdekis12, ricciardelli12}, a recently
extended version of the MILES \citep{vazdekis10} and CaT models \citep{vazdekis03}.
The MIUSCAT models are a library of single stellar population spectral
energy distribution 
(SED), covering the spectral range $3500-9500$ \AA\ at moderately high
resolution (FWHM $2.5$ \AA). The SEDs have been built by combining
three empirical stellar libraries: the MILES library \citep{Sanchez06}
covering the
range $\lambda\lambda\, 3525-7500$ \AA, the CaT library
\citep{Cenarro01}  in the range $\lambda\lambda 8350-9020$ \AA\
and the Indo-U.S library 
\citep{Valdes04},  used to fill-in the gap between the MILES and CaT
libraries and to extend blueward and redward the
spectral coverage of the SEDs.

Each stellar particle in the simulation is treated as a 
simple stellar population
(SSP), formed at a given time with a specific mass, metallicity and a
Salpeter IMF. Hence, we can assign a spectrum to each particle by
choosing the MIUSCAT model having age and metallicity closest to those of the
particle. Finally, to derive fluxes and magnitudes we make use of the SDSS passbands and the AB system.

\section{Finding and shaping the virtual galaxies}\label{shaping}

\subsection{The halo finding process}

The outcome of  our simulation is a complete  description of the three
components included  in the simulation,  namely, gas, dark  matter and
stars.  In  order to  analyse and characterise  the properties  of the
galaxies at the  different outputs, we identify the  galaxies by means
of an adaptive  friends of friends algorithm applied  only to the star
particles. In  the practical implementation  of our finder,  we linked
star  particles  using an  iterative  process  starting  from a  large
linking  length of  $\sim 100  \, Kpc$  and reducing  it until  a limit
length of $\sim 3 \, Kpc$.  Once all the particles belonging to a
given 
halo are  identified by the iterative linking  process, they undergone
an extra  process to check  whether they are gravitationally  bound to
the systems. Those unbounded particles are drop off the list of members
of such galaxy.  We experimented intensively with the different choices of
linking lengths, being the previous one the most stable.
Finally, we build the merger tree of each halo by looking at its progenitors 
 at the previous snapshots of the simulation. We consider 
31 snapshots between z=4 and z=0, with a typical time interval of 0.5
Gyr. The main progenitor is defined as that halo that contributes the most to
the stellar mass of the final halo.
When more than one
progenitor are present, we consider that a merger has taken place
when the mass ratio between the main progenitor and the
satellite\footnote{We define satellites as all those progenitor galaxies
  that are not the main progenitor} is higher than 0.025. Therefore,  the accretion of very small
haloes is not considered as a merger event.
We do not carry out any special treatment to identify satellites, as we are only 
interested in tracking the smaller galaxies  that eventually would merger with the 
main progenitor of the final galaxy.

The result of the halo finding process is a complete sample of all the
galaxy-like  objects in  our simulation  at the  different redshifts.
Every galaxy  is perfectly defined and all  its properties determined,
therefore,  the  generated  catalogue  can  be  used  to  explore  the
properties of the galaxies and to compare with the observational plane.
 
The analysis of the virtual Universe generated by our simulation by means of the 
previously described halo finder, produces a sample of galaxies spreading over a huge 
range of masses and sizes. In the present work, we focus our  analysis on the more
massive galaxies in the sample, those with zero-redshift stellar masses
$M_*>10^{11}\, M_\odot$. The total number of galaxies in our sample satisfying 
such mass condition is 33.
Galaxies in the process of merging are excluded from the
analysis, as their dynamical and morphological state are far from
being relaxed, hence difficult to characterize. We have therefore 
restricted the sample to galaxies that have not undergone
very recent merger events  and that are located in the higher resolution
grid. This leaves us with a sample of  21 galaxies more massive than $10^{11} M_{\odot}$.

\subsection{Baryon conversion efficiency}

\begin{figure*}
\includegraphics[width=16 cm]{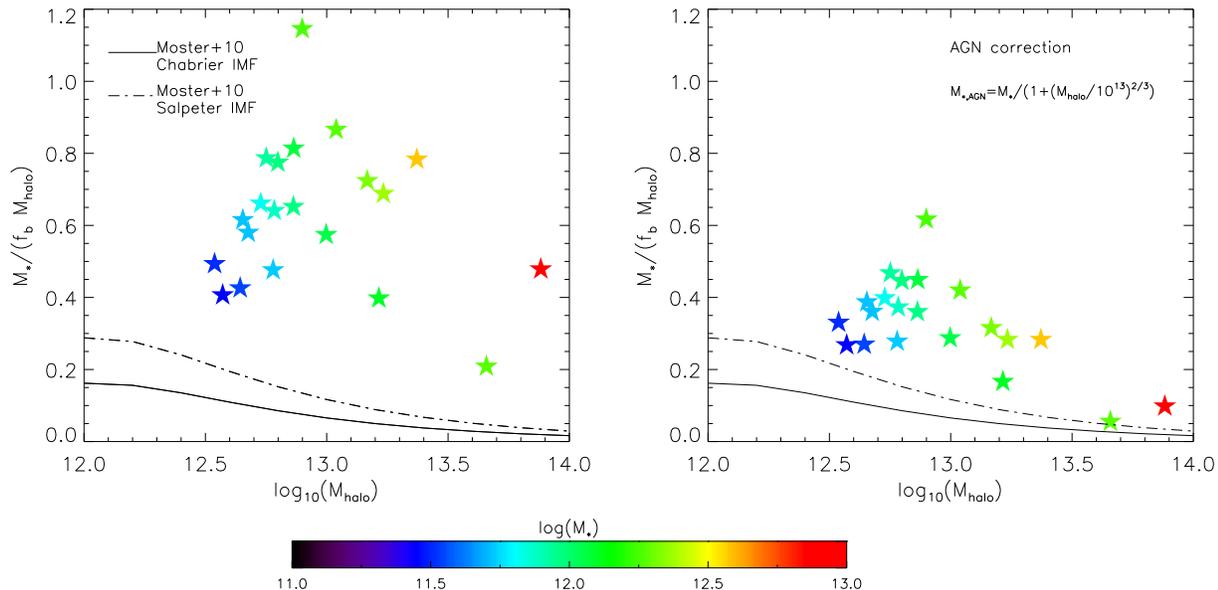}
\caption{Stellar to halo mass ratio as a function of halo mass for the 21 massive galaxies in our
  sample. Points are coloured according to the palette representing the galaxy stellar mass, 
  with the reddest points being the most massive galaxies. We also show the
  results of \citet{Moster10} from abundance matching tecniques for a
  Chabrier IMF (solid line) and Salpeter IMF (dashed line). In the
  left-hand panel we use the original stellar masses  from the friend
  of friends  halo finder,
  while in the right-hand panel we show the effect on the stellar
  masses that we would expect from the inclusion of AGN feedback. See text
  for further details.  }
\label{shm}
\end{figure*}

Since in our simulation we do not include AGN feedback, we expect our
stellar masses to be significantly biased towards higher masses.
To quantify  the efficiency of conversion of baryons into stars, 
in Figure \ref{shm} we show the 
baryonic conversion efficiency: $f_{conv}=M_{*}/(f_bM_{halo})$, where
$f_b=\Omega_b/\Omega_m$ is the cosmic baryon fraction, as a function of
halo mass.
Halo masses have been determined by means of the code 
ASOHF (Adaptive Spherical Overdensity Halo Finder,
\citealt{Planelles10}), applied to the dark matter particles. 
As a comparison, we also show the observational results from abundance matching
tecniques from \citet{Moster10}.
The original stellar masses of \citet{Moster10} have been calculated
assuming a Chabrier IMF (solid line in Figure \ref{shm}). Hence, to compare to our Salpeter-based stellar
masses we translate them according to \cite{cimatti08}:
$log(M_{Salpeter})=log(M_{Chabrier})+0.25 dex$ (dashed line).

\begin{figure}
\includegraphics[width=8 cm]{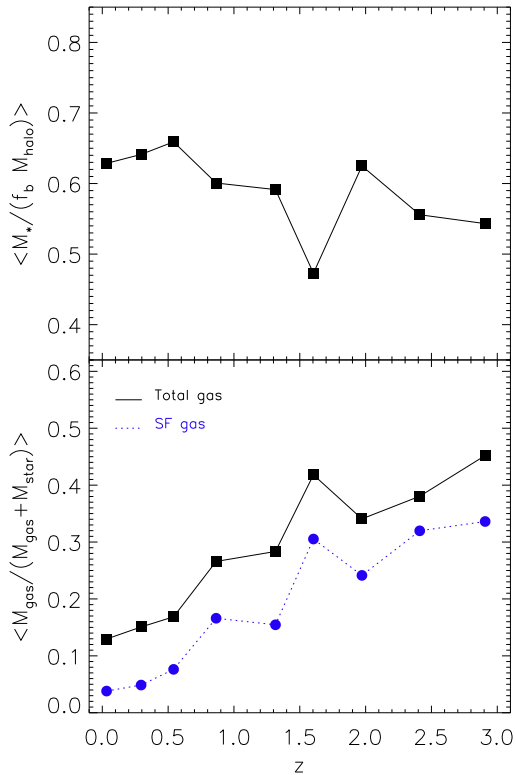}
\caption{Redshift evolution of the average baryonic conversion
  efficiency (upper panel) and of the gas fraction (lower panel). In
  the lower panel, the black line indicates the gas fraction when the
  total gas mass is considered and the blue line stays for the gas
  fraction when only the star forming gas is taken into account. } 
\label{frac_z}
\end{figure} 

Our galaxies appear overly massive for the halo in which they
live. The mean conversion efficiency for the whole sample is
$f_{conv}=0.63$, whereas the observations predict, for the same range of halo
masses, $f_{conv} \sim 0.15$.
However, since in the simulation we do not take into account AGN feedback, we
expect massive galaxies to be extremely efficient in the formation of
stars. As suggested by \citet{Cen11}, the effect of AGN feedback is to
alter the stellar masses by a factor:
$f_{AGN}=1/({1 + (M_{halo}/10^{13})^{2/3}})$. Hence, in the
right-hand panel of Figure \ref{shm} we show the baryonic conversion
efficiency using the stellar masses corrected by this
factor. The mean conversion efficiency significantly reduces to
$f_{conv}=0.33$, though a discrepancy of a factor 2-3 persists.
As shown in Figure \ref{frac_z}, this over-efficient star formation
occurs at all epoch at a quite constant rate. Thus, the stellar mass
assembly of the massive galaxies appears regulated only by the
dark matter merger rate. 
 In the lower panel of Figure \ref{frac_z}, we show the redshift evolution of the gas
fraction for  the total gas mass, 
$M_{gas}/(M_{gas}+M_{stars})$, and the star forming gas,
$M_{SFgas}/(M_{SFgas}+M_{stars})$. The mass of the star forming gas is
given by summing up the mass of the halo cells with $T  <
2\times10^4\,    K$   and   $\rho    >
\rho_{_{th}}=10^{-25}\,  g\,  cm^{-3}$ (see Section
\ref{starformation}). The gas fraction shows a strong evolution with
cosmic time, indicating that the progenitors of the massive present-day
galaxies at high redshift are gas-dominated systems. At low redshift,
massive galaxies contain a small but not negligible amount of gas,
with mean values of the  gas fraction of  0.13
and 0.04, for the total gas and the star forming gas, respectively. 

It is already known that hydrodynamical simulations lacking AGN feedback tend to produce too
many stars, leading to too massive galaxies. The problem is common to
both smoothed particle hydrodynamics (SPH) and grid codes, but it appears more
pronounced in the latter. 
For instance, in \citet{oser10} (SPH code GADGET-2), the overproduction of baryons
is less serious, with a conversion efficiency of the
order of $f_{conv} \sim 0.2$ for the same range of halo masses probed
here. 
On the other hand, simulations using AMR codes tend to produce higher 
conversion efficiencies. Indeed, \citet{lackner12} and \citet{Dubois13}
(in the simulation without AGN feedback) found a conversion efficiency
very similar to ours for a similar halo mass range. 
In the \citet{Dubois13} simulation with AGN feedback, the
baryonic conversion efficiency reduces down to a value of 0.1, leading
to stellar masses in a better agreement with the expectations from 
abundance matching techniques, although in the most-massive haloes they
remain too massive. It is important to notice that if the AGN feedback were as strong as suggested by
\citet{Dubois13}, the corrections applied to our stellar masses in
Figure \ref{shm} would
be then underestimated. 
Indeed, the values of the star forming gas fraction shown in Figure
\ref{frac_z} are in good agreement with those found by \citet{Dubois13} for the no AGN
case. As shown by those authors, the inclusion of a radio mode feedback
is particularly effective in reducing the amount of star forming gas at
low redshift, thus suppressing the late star formation and bringing
down the discrepancy with observations. 

Several reasons can  contribute to explain the  differences found among different codes.
Spatial and mass resolution can in principle affect the results on the
star formation efficiency, with low-resolution simulations expected to
give higher late star formation.
We have a mass resolution comparable with that of
\citet{oser10}, but a coarser spatial resolution, although the
smoothing length in SPH cannot be directly compared to the cell size
of AMR codes. 
We expect that an improved resolution can help in reducing the
problem, but  we should note that 
AMR simulations using higher spatial
resolution than that used here \citep{lackner12, Dubois13}, produce the same amount of
over-production of stars. Hence, we do not expect that improved
resolution would be the only solution.
The star formation efficiency can also affect the final results,
because  higher efficiencies produce higher star formation at high
redshift and a lower level of late star formation.
However, \citet{Dubois13} have shown that an increased star  formation
efficiency does not lead to significant changes in the final stellar mass.
We also note that  the star formation efficiency used in this work is
similar to that used in \citet{oser10}.

Hence, we are prone to think that in the absence of AGN feedback,
the amount of gas converted into stars is primarily regulated by the
cooling efficiency.
Comparisons between SPH and grid codes (sharing  same physics and
same initial conditions) have led to the conclusion that SPH codes
produce less efficient cooling rate. This naturally translates in lower
levels of star formation rate and less massive galaxies 
\citep{Keres12, Scannapieco12}. 
The reason for that would be in the different treatment of shocks and
gas instabilities among the two numerical techniques.
As demonstrated by \citet{Agertz07}, gas instabilities are not
correctly solved in SPH codes. 
The different treatment of hydrodynamical instabilities in the
  two numerical techniques can lead to importance differences in the dissipation
  heating. \citet{vogelsberger12} have shown that  SPH codes produce a
  significant heating in the outer part of the haloes, where the
  cooling radius is expected to lie. This produces an overall higher temperature
  and a weaker cooling efficiency  that can
  explain the lower level of star formation in these simulations (see also \citealt{Keres12}). However, this dissipation
  in SPH is likely to be of spurious nature and ultimately caused  by
  errors in the estimation of pressure gradients. This different behavior of SPH and grid techniques could be than a
potential source of discrepancy among the \citet{oser10} findings and
our results.

\subsection{Two dimensional maps}

A proper characterization of the morphological structure of the
simulated galaxies requires to treat them as close as
possible to real objects. 
To this aim, the 3D structure of a galaxy
has been converted in a 2D map by projecting its volume of star
particles onto a plane.   
To overcome the resolution limit we choose a pixel size in the 2D
image equal to the 
spatial resolution of the MASCLET level where the galaxy is
located. Since we have selected only galaxies in the highest level of
refinement ($l=7$), the pixel size of the 2D map is 2.7 kpc for all the
galaxies in the sample, that roughly corresponds to $R_e/4$ for the
smallest galaxies and $R_e/8$ for the largest ones.
The galaxies have  been projected along each one of
the coordinate axes of the computational box.
In this way we create three different 2D maps for every 
simulated galaxy.  These 2D projections have been used to get the 1D
profiles described in the next section.

\begin{figure*}
\includegraphics[width=16 cm]{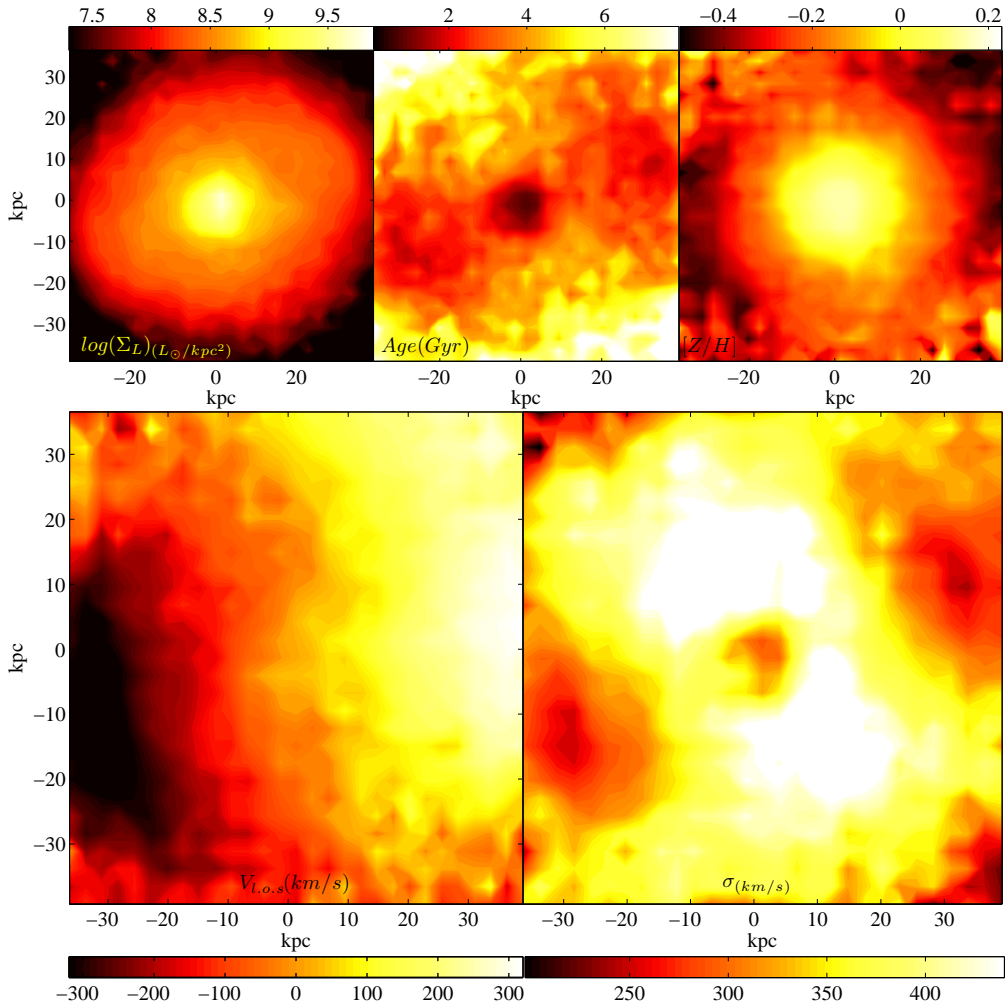}
\caption{2D map of one of the most massive galaxies  in our sample
  ($M_* \sim 3.3\times 10^{12}\, M_\odot$ or $M_*=1.5\times 10^{12}\,
  M_{\odot}$ after AGN correction, $R_e= 14.2 \, Kpc$) . Top panels represent, from left to right: logarithm of 
the surface density mass, age of the stellar population, and metallicity in solar units. Bottom panels show the 
line-of-sight velocity (left) and the velocity dispersion (right). According with the three classification criteria described in 
Sec. 4, this galaxy is just at the transition between the fast and slow rotator types from its dynamics.  Its history labels it as 
a merger galaxy, whereas its morphology indicates that it is an elliptical galaxy.}
\label{2Dmap}
\end{figure*}

To create the artificial images, the total flux in every pixel is computed by summing up the flux of each
SSP weighted according to its mass. 
To limit the galaxies to their visible part we only
consider pixels whose surface brightness (SB) in the r-band is
brighter than  25 mag/arcsec$^2$,
which corresponds to typical  limiting SB in the observations.  
An example of artificial images for a representative galaxy in the
sample is shown in Figure \ref{2Dmap}.
Artificial images have been used to derive the galaxy structural
parameters through the two-dimensional fitting code GALFIT \citep{peng02}.
The light distribution of the simulated galaxies has been modelled
with a S\'ersic profile \citep{sersic68}, deriving the S\'ersic index $n$, the semi-major effective
radius $R_e$ and the axial ratio $b/a$. Although in the following we
make use of the S\'ersic fits derived by means of 1D profiles, the 
axial ratio assigned to each galaxy is the one derived with GALFIT.

Moreover, the use of two-dimensional grids  allows us  to measure the velocity
dispersion along the line-of-sight (LOS) in a consistent way.
Once we have computed the mean velocity of the stellar particles in
each pixel, the velocity dispersion of the particles in the cell is
estimated through the square deviations of particle velocities from
this mean value (see the Appendix). 
 Therefore, the velocity dispersion in the individual pixels does not
 need any further correction for galaxy rotation.

In Figure \ref{2Dmap} we show, as an example, the  2D maps of the relevant quantities analyzed  
in the present work for one of the most massive  galaxies of the sample.
 
\section{Results}\label{results}

We study the sample of massive galaxies and their more relevant features attending to 
three wide criteria: (i) their dynamical properties, (ii) their evolutionary history, and (iii) their morphologies.

As a first step, we classify the galaxies in our sample according to the dynamics. To do so, we look at 
a quantity widely used in the literature (e.g. \cite{emsellem11}): the ratio of the rotational velocity to
the dispersion velocity ($V/\sigma$). This quantity allows us to 
split the sample into two groups, the slow rotators and the fast
rotator objects. 
   
The second criteria that we use to study the galaxies in our sample is their evolutionary history. Thus, according to their evolution, 
we separate the galaxies in two categories: those galaxies which have
suffered at least a merger event, 
and those other galaxies that have a quiet evolution without any
mergers recognized by the halo finder. Thus, the merger sample also
includes galaxies having experienced only minor mergers.
    
Finally, the third criteria used to sort the sample is the morphology. The basic methodology consist in fitting the light profile of 
each object by the corresponding S\'ersic profile and obtain the S\'ersic index ($n$).

\begin{figure}
\includegraphics[angle=0,width=\columnwidth]{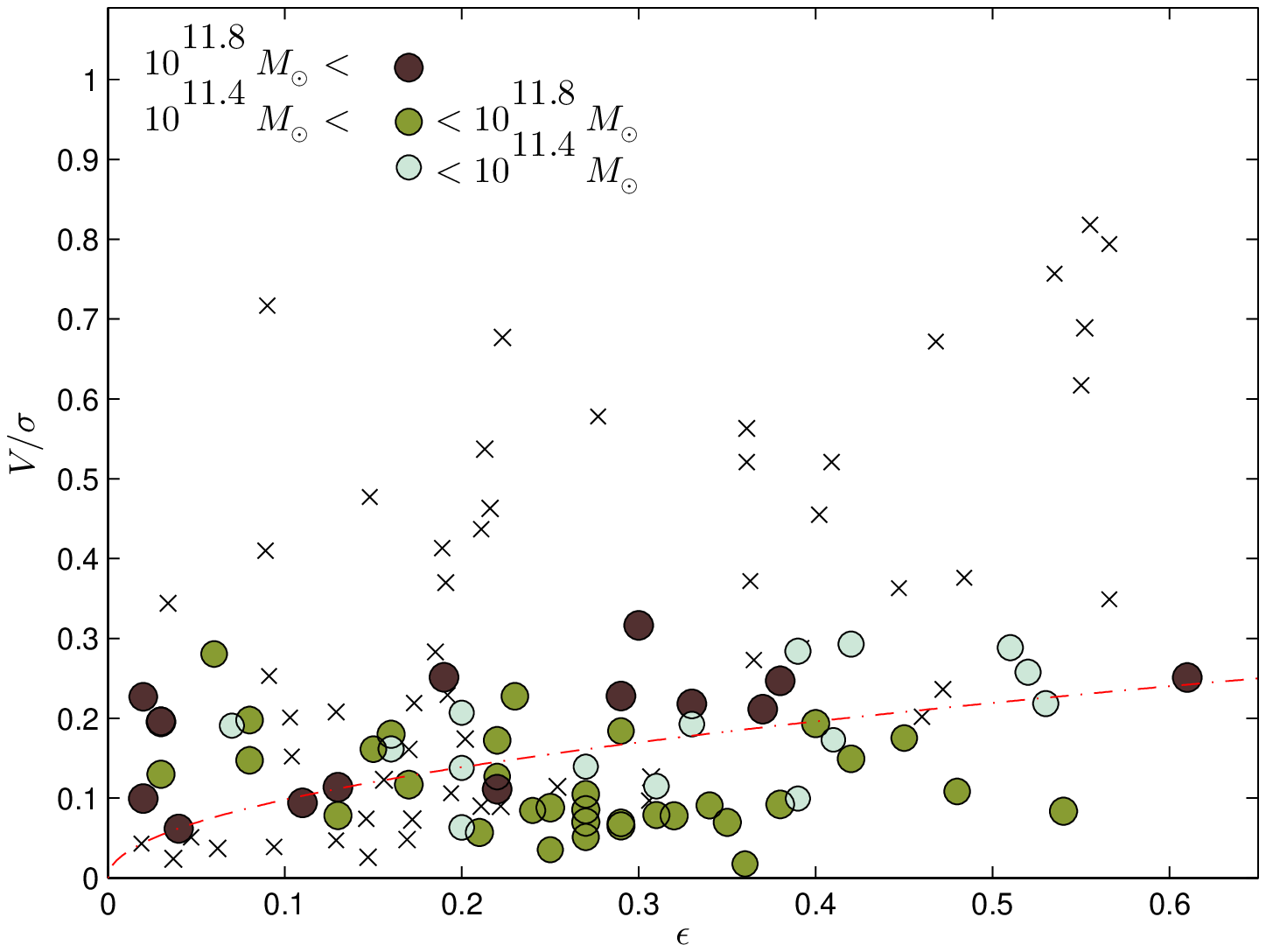}\\
\includegraphics[angle=0,width=\columnwidth]{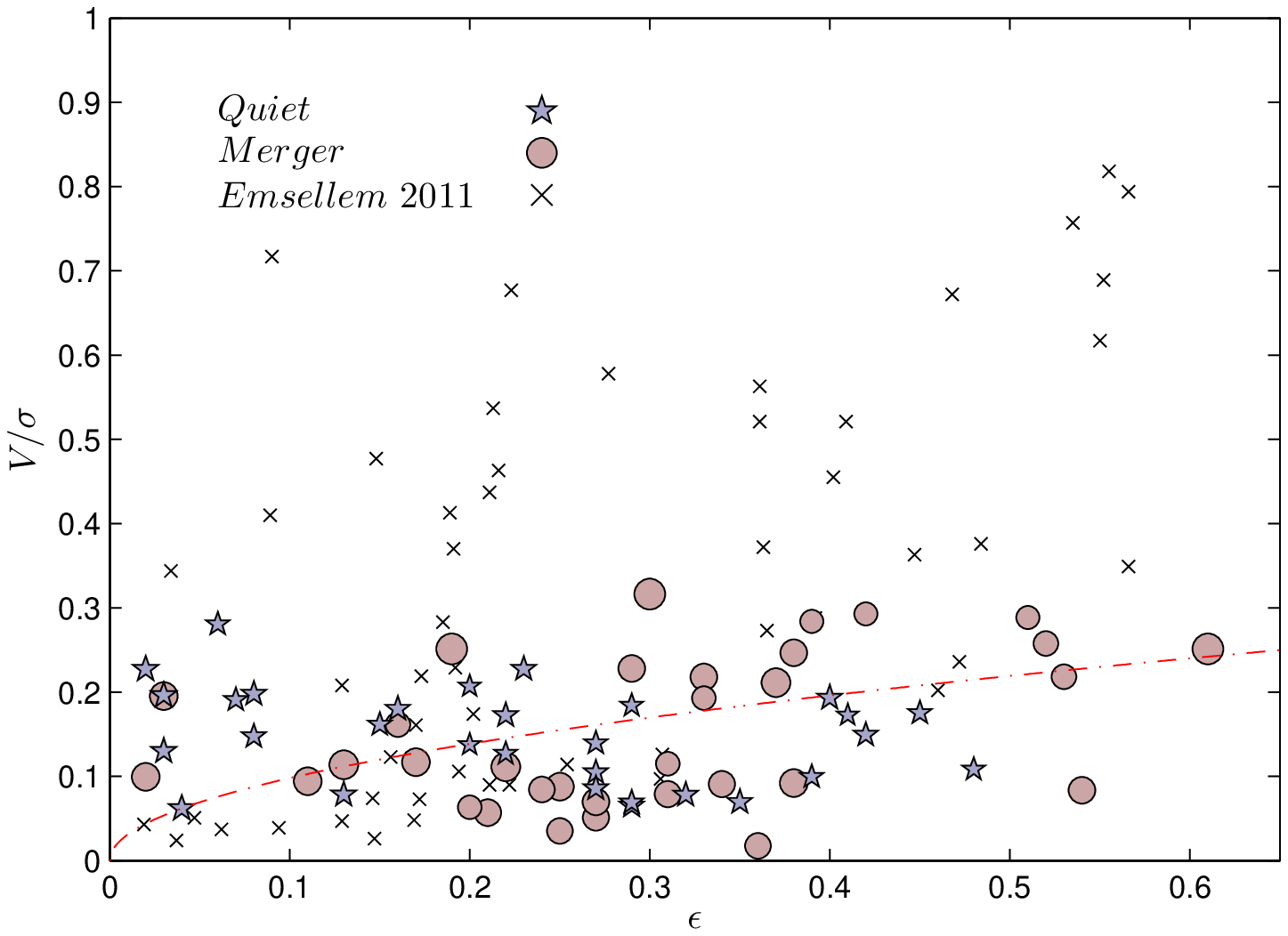}\\
\includegraphics[angle=0,width=\columnwidth]{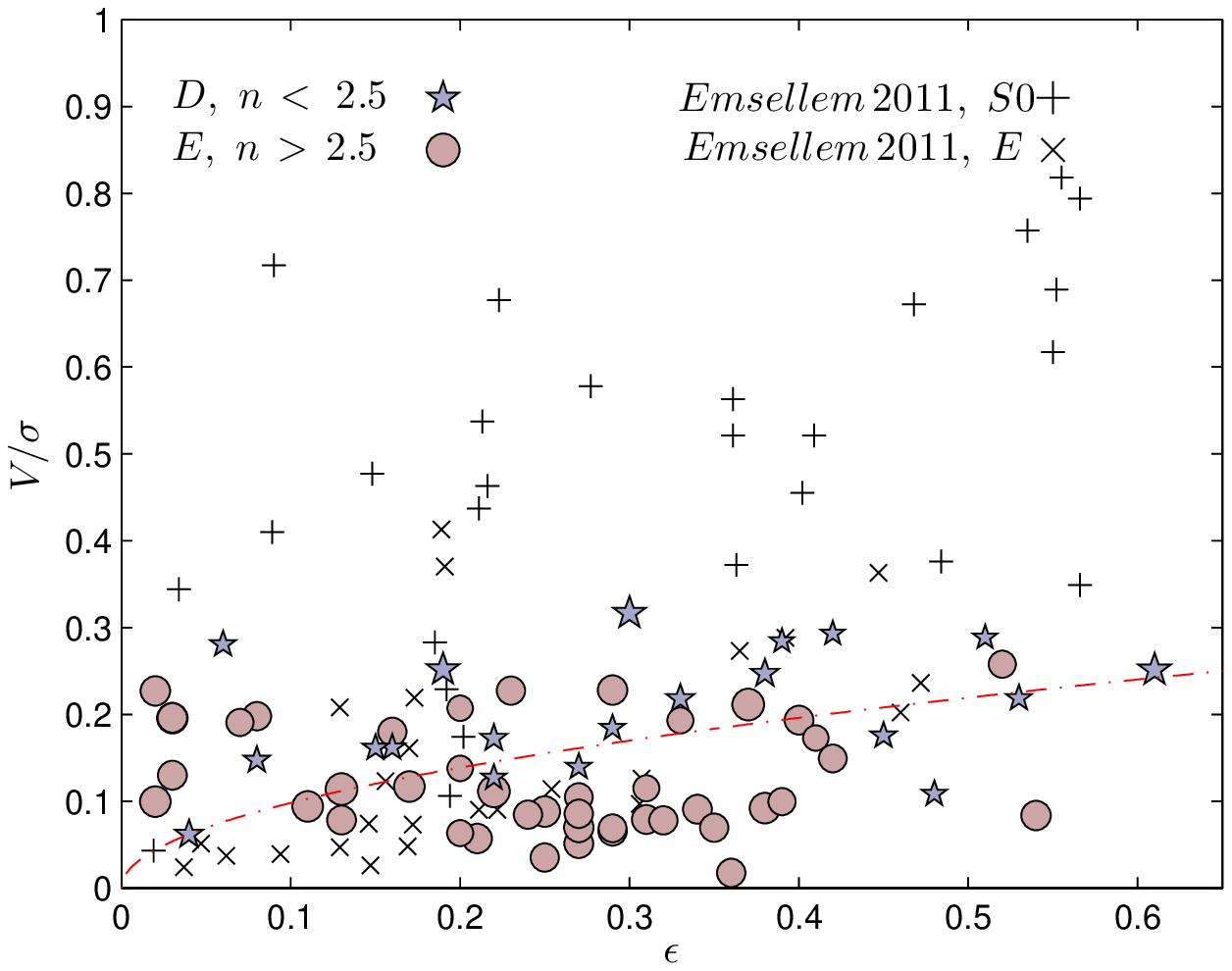}
\caption{$V/\sigma$ vs ellipticity diagram for our simulated galaxies
  (color symbols). For the sake of comparison, crosses indicate the position in
  the diagram of the most massive galaxies in the ATLAS sample of
  \citet{emsellem11}, whereas the red dot-dashed line is
  $V/\sigma=0.3\sqrt{\epsilon}$, used to discriminate slow and fast
  rotators. In the top panel galaxies are
  colored according to their stellar mass, and in the middle (bottom) panel  they are separated
  according to the merger history (morphology). In the bottom panel,
  the observational data points are also separated into early-type
  galaxies and lenticulars. }
\label{v_sig_obs}
\end{figure}

In order to make our results more consistent with the
  observational data, and according with the previous discussion in
  Sec. 3.2, all the galaxy masses in the sample are corrected
  following the prescription by \cite{Cen11}. This post-processing of
  the galaxy masses  allows us to modify our results for the stellar masses in a way that
  tries to mimic the effects of the AGN feedback.
  We have also to remark that  the rest of the galaxy
  properties that could change due to AGN feedback, like stellar populations and
  velocity dispersion, are kept unchanged. 

\subsection{One dimensional profiles}

One dimensional (1D) profiles are useful tools to compare the results of the 
simulations with observational data. To produce 1D profiles as similar as possible to 
the ones produced by observations, we perform the following
procedure. 

In each of the three projections we identify the center of the 2D map. 
This is done in a two steps process. First,  the centre is computed as
the centre of luminosity or mass, depending on  whether we consider mass or light
weighted profiles, of all the star particles forming the
galaxy. Then, the effective radius ($R_e$) is defined as the half
light (mass) radius. As for the total luminosity (mass) used to
determine the effective radius, 
we only consider the luminosity (mass) in cells whose SB is brighter than 25 mag/arcsec$^2$.
A finer determination of the centre of the galaxy is given by
computing  the centre of luminosity (mass)  only with the particles inside  
the half light or mass radius. Finally, starting from the centre, the radial 1D profile for the considered quantity is obtained by 
averaging in circular shells whose width is such that contains one per cent of the  total mass of the galaxy for mass weighted profiles or 
one per cent of the total flux in the case of light-weighted
profiles. We choose  this particular binning in order to be able to reach the
external part of the galaxies, while at the same time achieve smooth profiles.

The previously described process allows us to produce 1D profiles of every galaxy in our sample. In particular, we analyse the 
following quantities: luminosity $\Sigma_L$, surface density $\Sigma_{M}$,
velocity dispersion $\sigma$, line of sight velocity $V$, age and metallicity $Z$.

Depending on the analysed criteria, galaxies are gathered in different groups. All the 1D profiles of the galaxies in the same group 
are averaged. The following figures shown in this Section present these 1D median profiles 
and the $25^{th}/75^{th}$ percentile as shaded regions. We use the
luminosity in the {\it r} band to light-weight the 1D profiles.

\subsubsection{Dynamics}

In numerical simulations, 
the definition of a quantity representing the velocity dispersion ($\sigma$)
 is somehow vague and can be defined in many different manners. Although, all those definitions can 
be self-consistent in the simulations, it becomes crucial how $\sigma$ is defined
when the virtual galaxies in the simulations are compared with observed ones. 
In order to get rid of this ambiguity, we have found the best definition of $\sigma$ in 
the simulations to compare with observations. In Appendix A, we  present a 
detailed explanation of the method used to estimate the velocity dispersion.
In brief, we find that a luminosity weighted mean of the square
deviations of the particle velocities (eq. \ref{sigma_lm}) shows a
close agreement with
the velocity dispersion derived from the spectral features.

As a proxy to characterize the rotational structure of our numerical
sample of galaxies, we have used the ratio of ordered to random motion
in a galaxy ($V/\sigma$), both projected along the line-of-sight (LOS)  
We have also checked that by separating the dynamical groups by means
of the $\lambda$ parameter, as defined in \citet{emsellem11}, the
results of this section do not change substantially.

In Figure \ref{v_sig_obs} we present the anisotropy diagram, relating
this parameter  to the observed
ellipticity ($\epsilon$).  The ellipticity is defined as
$\epsilon=1-b/a$, where $b/a$ is the galaxy axial ratio, measured by
means of GALFIT. 
Following the approach used in integral-field studies
  \citep{cappellari07, emsellem11}, $V/\sigma$ is defined by:
\begin{equation}
\Big( \frac{V}{\sigma} \Big)^2=\frac{\sum\limits_{i=1}^{N}{L_iV_i^2}}{\sum\limits_{i=1}^{N}{L_i\sigma_i^2}}
\end{equation}
 where the summation extends over  the stellar particles lying within
the effective radius, $L_i$ is the particle luminosity
and $V_i$ and $\sigma_i$ are the LOS velocity and velocity
dispersion associated to the star particle.
In the upper panel of Figure \ref{v_sig_obs}, our data are shown as circles colored according to their 
masses. 
For each galaxy, all the three projections are shown. 
Overplotted to our data, in Figure \ref{v_sig_obs} we show the position
in the diagram of the most massive galaxies (selected to have magnitude
in the K-band brighter than $-23.8$ mag) in the ATLAS sample
\citep{emsellem11}. According to these authors, fast and slow
rotators can be well separated in the  $V/\sigma$-$\epsilon$ plane
using as threshold: $V/\sigma=0.3\sqrt{\epsilon}$, that is shown as a
magenta dot-dashed line in Figure \ref{v_sig_obs}.
Therefore in the following we define slow-rotators (S) the galaxies lying
below this line, that constitute the 54\% of the total sample and
fast-rotators (F) the remaining galaxies (46\%). 

The simulated galaxies span a wide range of rotational properties, in
agreement with the distribution observed for the bright ATLAS
galaxies. 
 However, the simulated galaxies do not reach the highest values
  of $V/\sigma$ displayed by the observed galaxies.
Indeed, the fraction of fast rotators in the ATLAS sub-sample is 69\%, 
higher than our value. 

Stellar mass does not seem to play a crucial role in 
discriminating among fast and slow rotators,
as the fraction of fast rotators (slow rotators) is similar in the three mass
intervals considered. Likewise, by dividing
the Emsellem sample in three bins according to their luminosity, we
get the same result, i.e. the fraction of F (S) does not change with
luminosity.

Although both quiet and merger galaxies can be fast rotators, the
highest values of $V/\sigma$ are reached by galaxies having undergone
mergers. Moreover, fast rotators are usually classified as late-type
galaxies (see bottom panel of Fig. \ref{v_sig_obs}), in agreement with
the Emsellem sample of fast rotators that show  S0 morphology.

\begin{figure*}
\includegraphics[width=15 cm]{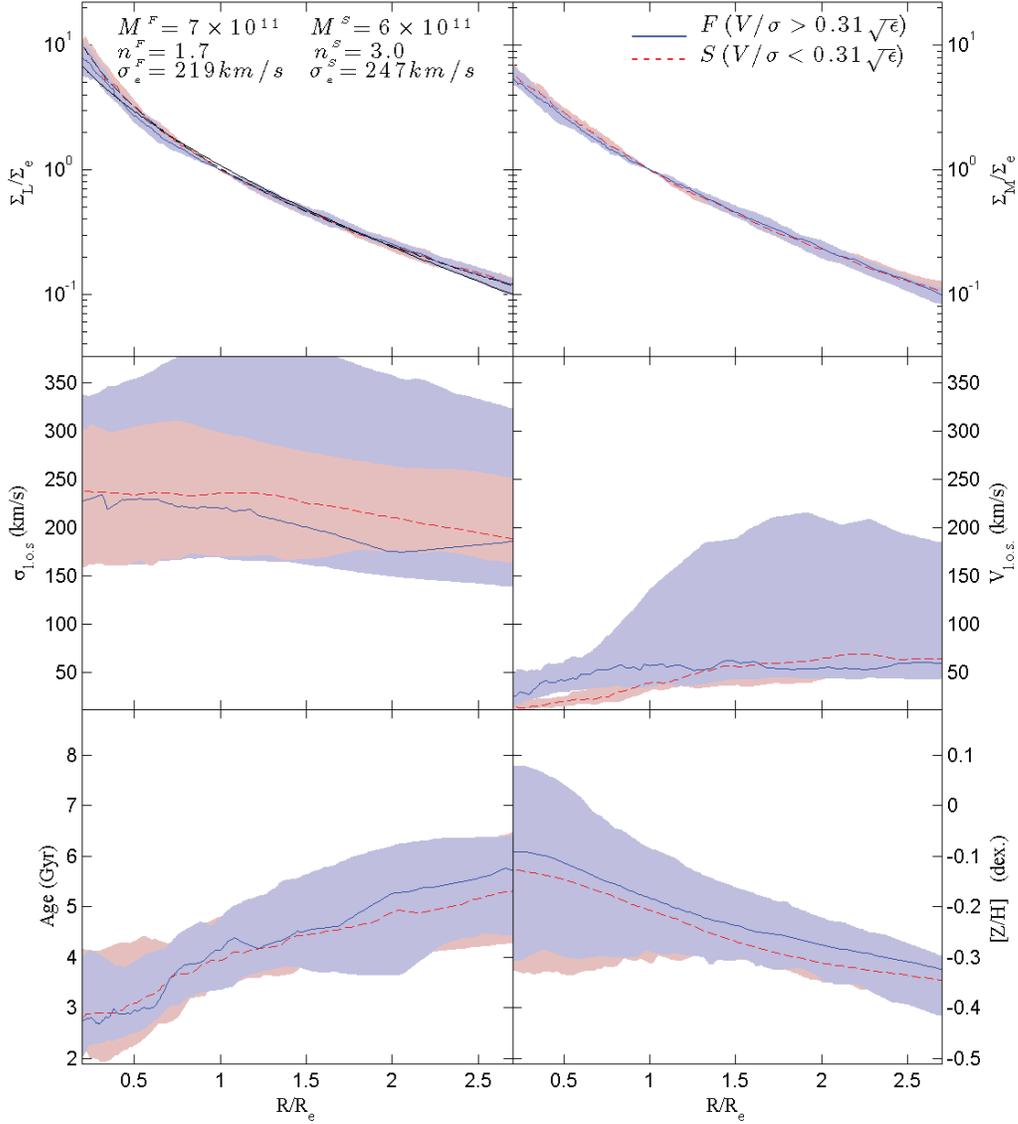}
\vspace{-0cm}
\caption{1D profiles for the galaxies in the sample grouped according
  to their dynamics: slow rotators (S)  with $V/\sigma<0.3\sqrt{\epsilon}$
and fast rotators (F) with $V/\sigma>0.3\sqrt{\epsilon}$. 
The lines represent the median of all the profiles of the
galaxies in every group and the shaded regions stand for the $25^{th}/75^{th}$
percentiles of the distribution. 
The panels represent: surface brightness (top left), surface density (top right),
velocity dispersion (middle left), line of sight velocity (middle
right),  luminosity-weighted age (bottom left), and
luminosity-weighted metallicity (bottom right). 
The blue solid line and the blue
shaded region correspond to the fast rotators (F), 
whereas the red dashed line and the shaded region stand for the slow
rotators (S).
In the top left panel the black dashed (solid) lines are the S\'ersic fits to the median light
profiles of the slow (fast) rotators. 
To average all the profiles corresponding to the galaxies in each group, the radial profile of each galaxy
is rescaled in units of its effective radius, $R_e$. The surface density and brightness profiles of each galaxy are normalized to their
values at  $R_e$.  All profiles displayed in this figure are light-weighted and they are calculated starting at $0.25R_e$.}
\label{dyn_vs}
\end{figure*}

Figure \ref{dyn_vs} presents the main properties of the 
galaxies in the numerical sample classified according to their 
values of $V/\sigma$. 
As all the 1D plots 
in this Section, lines represent the median of all the profiles of the 
galaxies in every group and the shaded regions stand for the $25^{th}/75^{th}$ 
percentile of the distribution. The panels represent: luminosity (top left), surface density (top right), 
dispersion velocity (middle left),  line of sight velocity
  (middle right), luminosity-weighted age (bottom left) and luminosity-weighted
metallicity (bottom right).
The blue solid line and the blue shaded region correspond to the fast 
rotator objects, whereas the red dashed line and the shaded region stand for the slow rotators.
In order to average all the profiles corresponding to the galaxies in each group, the radial profile of each galaxy 
is rescaled to its effective radius, $R_e$.

The fast rotator (slow rotator) galaxies
have a median stellar mass of $M_{_F} \sim 7\times 10^{11}\, M_{\odot}$ 
($M_{_S} \sim 6 \times 10^{11}\, M_{\odot}$)  
and a velocity dispersion at the effective radius of $\sigma_{_F} \sim 219 \, km/s$ ($\sigma_{_S} \sim 247 \, km/s$).
These previous stellar masses are the values taking into account the AGN correction. For the sake of completeness, the stellar masses without correction are  $M_{_F} \sim 9\times 10^{11}\, M_{\odot}$ 
 and $M_{_S} \sim 1 \times 10^{12}\, M_{\odot}$.
Although the light and mass profiles of the two galaxy groups look
like quite similar, there is a tendency for the slow rotators to have
higher S\'ersic indices ($n_{S} \sim 3.0$) than galaxies with higher
rotational support ($n_{F} \sim 1.7 $). 
As expected, there are differences in the line of sight
velocity, with the fast rotators presenting higher values of $V$
within the effective radius and
therefore higher rotation.
The velocity dispersion within the effective radius is very similar in 
the two groups, but at larger radii the velocity dispersion of the
fast rotators decays more rapidly. 

The analysis of the stellar populations of the two groups is summarized in the two bottom panels of Figure \ref{dyn_vs}. 
There are significant age gradients in both groups. 
The general trend is a positive age gradient,   meaning younger 
  ages at the center that translates into steeper age gradients
within the effective radius for both groups of galaxies.
Our findings of positive age gradients is
consistent with the observational result of \citet{labarbera12} and 
 \citet{ coccato10}, that have determined radial profiles of early-type galaxies out to
 large galactocentric distances.
In the case of the metallicity, both groups show negative gradients, with the slow
rotators having a slightly lower metallicity. 
Despite the trends found in the age and metallicity profiles, we must warn on the values of the ages, which are too low, and the 
the metallicities at the center which do not reach solar values as it would be more consistent with observational data.

\subsubsection{History}

In an hierarchical scenario for the formation of cosmic structures,
mergers have a crucial role as they 
sculpt the main features of the galaxies and galaxy clusters. 
Therefore, 
the evolutionary history of galaxies is an important key to understand 
their actual properties \citep{mihos94, cox06, hopkins09}.
Within this paradigm, there must be 
substantial differences between those objects which have had a 
relatively quiet life (with no important merger events) and those 
involved in major merger processes. 

Following a similar procedure than in subsection 3.1.1, 
we group the galaxies in the sample separating those which have 
undergone mergers (M, 52\%) and those that have had a quiet evolution
(Q, 48\%).  
We assume that a merger has token place when the mass ratio
  between the satellite  and the main progenitor is larger than 0.025.
Among the mergers, 
only few of them (21\%) can be considered a major merger (those events with a mass ratio between the objects involved in 
the merger larger than $\sim 0.3$). Thus, the majority of merger in the simulation (79\%) are classified as minor ones,  with an average mass ratio of the galaxies involved around $0.17$.
The M (Q) type galaxies 
have a median stellar mass -- including the AGN feedback correction -- of $M_{_M} \sim 8\times 10^{11}\, M_{\odot}$ 
($M_{_Q} \sim 6\times 10^{11}\, M_{\odot}$), a S\'ersic index fitted from the median light profile 
$n_{_M} \sim 1.7$ ($n_{_Q} \sim 2.8$),
and a velocity dispersion at the effective radius of $\sigma_{_M} \sim 256 \, km/s$ ($\sigma_{_Q} \sim 223 \, km/s$). 
For the sake of completeness, the uncorrected stellar masses are $M_{_M} \sim 1.6\times 10^{12}\, M_{\odot}$ 
 and $M_{_Q} \sim 1.1\times 10^{12}\, M_{\odot}$.

We also characterize the kind of merger occurred by
looking at the SFR at the time of merger. 
A dissipational merger can be defined as that merger triggering a strong
starburst.
Several studies of star forming galaxies indicate the existence of a
main-sequence in the SFR-$M_*$ plane where most of the star forming galaxies
lie \citep{daddi07, noeske07, rodighiero11, whitaker12}. 
A deviation from this sequence towards high SFR indicates the
occurrence of  a starburst. Taking into account the redshift dependence of
the SFR-$M_*$ normalization, we adopt the criterion defined 
in \citet{whitaker12} to 
establish the presence of the starburst during a merger event:
\begin{equation}
log(SFR) > \alpha(z)(logM_{*}-10.5)+\beta(z)+0.34
\end{equation}
 where $\alpha(z)=0.70 - 0.13z$ and 
$\beta(z) = 0.38 + 1.14z - 0.19z^2$, 
$M_*$ is the mass of the galaxy after the merger and the value of 0.34 dex represents the  $1\sigma$ scatter used to
 define the outliers. 
Concerning the SFR, we use the mass of gas converted in stars during the last time-snapshot 
 before the merger is identified. 
We find that 10 out of 11 merging galaxies have undergone
 a dissipational merger, while just in one case the merger
 can be considered dry. 

\begin{figure*}
\includegraphics[width=15 cm]{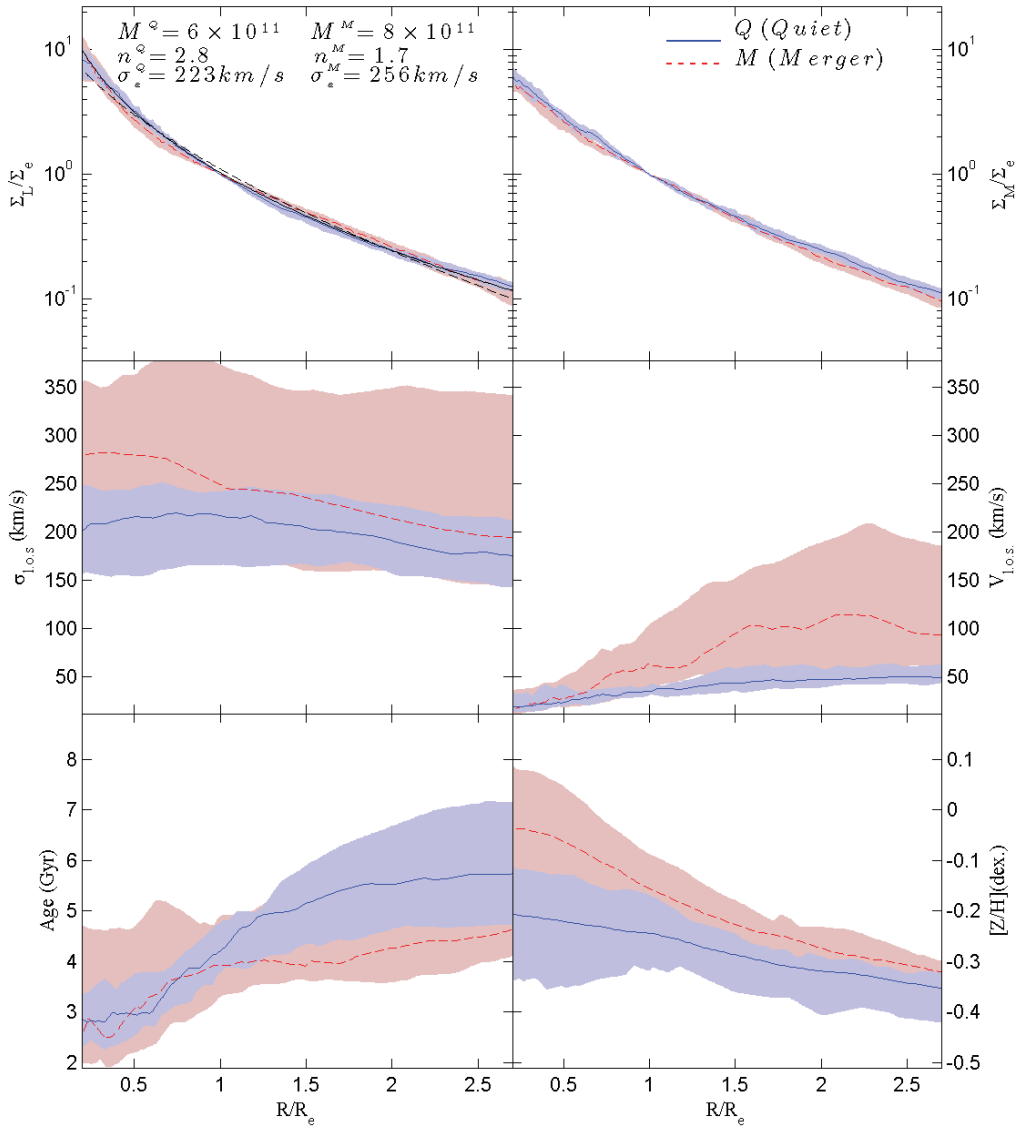}
\vspace{0cm}
\caption{The same as Figure \ref{dyn_vs} when galaxies are split according
  to their merging history. Red lines and shaded regions stand for
  galaxies having experienced a merger (M), while blue corresponds to galaxies
  with a quiet life (Q).
}
\label{history}
\end{figure*}

In Figure \ref{history}, we present -- in a similar way than in Fig. \ref{dyn_vs} -- the median 
profiles for the studied quantities for the 
two groups of galaxies: $\Sigma_L$, $\Sigma_M$, $\sigma$, $V_{_{LOS}}$, age, and metallicity. 

According to the expectations, the evolutionary history of galaxies
seems  
to be an important factor determining their features. 
Although light and density profiles in  Fig. \ref{history}  are very similar, as it happened in the 
study based on the dynamics,
in all the other analysed quantities the differences are notable. Thus, 
galaxies which have undergone mergers (M) exhibit larger 
velocity dispersion at all radii and a strong velocity dispersion
profile, which is almost flat within the effective radius and decays 
abruptly outside. 
The line of sight velocity is also higher for the 
M galaxies except at the inner part ($R<0.5R_e$), where 
both types of galaxies show similar values. These low values indicate that the very central parts do not present an 
important rotation regardless the evolutionary history of the galaxy. 
At outer radii ($R>R_e$), the rotational support of the M galaxies
increases significantly.  
Numerical simulations of mergers (Cox et al. 2006) have indeed shown
the relevance of the merger-induced dissipation in the formation of fast
rotators. Since the majority of the mergers suffered by our simulated
galaxies involves a significant fraction of SFR, our results are
consistent  with those findings.

The galaxies without 
mergers (Q) are older at all radii and less metallic out to the
external regions. 

The younger nature of the merger galaxies is not a surprising result given the fact that the
majority of mergers have triggered a starburst rejuvenating the
stellar populations. At the same time, the star formation occurring in
an already enriched medium can produce metal-rich stars, explaining the
metallicity trend.
The most remarkable feature in the stellar population profiles is the
steepness of the metallicity gradient of the M galaxies. 
A number of studies have explored the role of dissipationless mergers
in the formation of stellar population gradients \citep{kobayashi04,
  dimatteo09},  finding  a flattening of the metallicity gradient  in
the remnants, mainly due to stellar mixing. However, if the merger is
dissipational, as in our galaxies, a rejuvenation of the stellar
contents in the central region is expected, establishing a steeper
metallicity gradient than that of the progenitor galaxy (see for
instance \citealt{hopkins09}).

We have verified that the trends found with the merging history are not just an effect of the different mass of the two groups or the nature of the merger events (major or minor). To do this,  we 
repeat the analysis using two subsamples having similar masses and considering only major mergers or minor mergers in the M class. We find no substantial changes in the trends found in Figs.~\ref{history}.  The only worth mentioning difference is the presence of a steeper metallicity gradients in the major merger subsample, whereas the subsample formed by the galaxies that had experimented minor mergers, exhibits metallicity gradients more similar to those of the Q galaxies.
Major mergers are thus the first cause for the steeper metallicity gradients observed in Fig. \ref{history}.
Therefore, the results on the dynamical profiles, the age and metallicity gradients are robust against the definition of the M/Q classes.

\subsubsection{Morphology}  

\begin{figure*}
\includegraphics[width=15 cm]{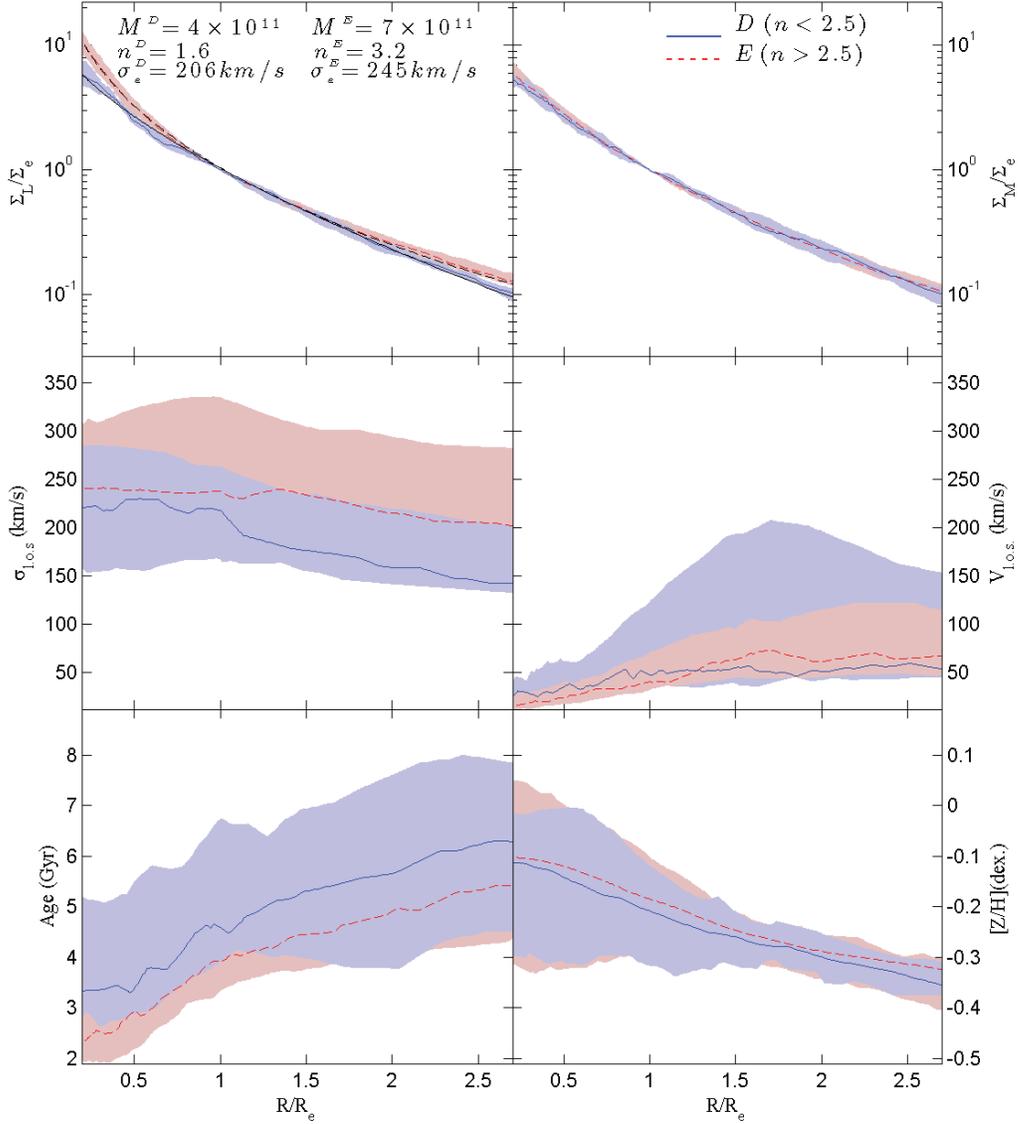}
\vspace{0cm}
\caption{
The same as Figure \ref{dyn_vs} when galaxies are split according
  to their morphology. Red lines and shaded regions stand for
  early-type galaxies (E), while blue stands for disk-like  galaxies (D).
}
\label{sersic}
\end{figure*}

To classify galaxies according to their morphology we fit the 1D
density profile with a S\'ersic model, deriving the S\'ersic index, $n$, 
the half-light radii, and the effective luminosity density. The radial
interval taken into account in the fits ranges from $0.25R_e$ up to $10R_e$.
The S\'ersic index is a common indicator used to classify the 
morphology of galaxies \citep{Blanton03, Ravindranath04}. Thus, galaxies with values for $n$ larger (smaller)
 that 2.5 are classified as elliptical (spirals). By using this classification, 
 Figure \ref{sersic} presents, as usual in this Section, the average properties of the galaxies in the sample
 classified according to their S\'ersic  index, $n$, into elliptical (E, 70\%)
 and disk-like (D, 30\%) galaxies.
As a general rule, the morphology classification derived by fitting the one-dimensional
light profiles well agrees with that determined by means of GALFIT,
although galaxies with intermediate S\'ersic indices ($n \sim 2.5$)
can be in a few cases misclassified.

In Figure \ref{sersic}, we present -- in a similar way than in Fig. \ref{dyn_vs} -- the median 
profiles for the studied quantities for the 
two groups of galaxies: $\Sigma_L$, $\Sigma_M$, $\sigma$, $V$, age, and metallicity. 
The mean stellar mass considering the AGN feedback correction for E (D) galaxies is 
$M_{_E} \sim 7\times 10^{11}\, M_{\odot}$ 
($M_{_D} \sim 4\times 10^{11}\, M_{\odot}$). If the AGN correction is not considered, the mean stellar masses 
are $M_{_E} \sim 1\times 10^{12}\, M_{\odot}$ and
$M_{_D} \sim 7\times 10^{11}\, M_{\odot}$. 

A mean S\'ersic index, $n$, for both morphology groups has been
derived by fitting the median profiles, obtaining 1.6 and 3.2 for the
disc and the elliptical galaxies, respectively.
These values allow to clearly differentiate between 
the two considered morphological types.

As expected, the light and density profiles are more different between the two types of galaxies than in any of 
the other classification previously used. 
The dispersion  velocity for both D and E galaxies is quite similar at
the inner parts and gets larger for elliptical  galaxies compared with
spiral galaxies as radial distance increases.

There are relevant age gradients in both categories of galaxies, with the 
internal part substantially younger than the outer parts. 
The elliptical galaxies are younger than disk galaxies at all radii.
Observed ellipticals are instead older  than galaxies with
a disk-like morphology. The fact that we are not able to reproduce
the observed trend can be partly explained with the fact that our galaxies are
not able to quench star formation efficiently. This leads to a tail of
star formation  at low redshift that contributes significantly to
rejuvenate the stellar population of both kind of galaxies. 
We argue that by improving the feedback scheme and the spatial
resolution, we would alter the star formation history of these
galaxies. 
Finally,
metallicities and metallicity gradients
are very similar for both subsamples.

\subsection{Star formation history: in-situ vs accreted}\label{insitu}

\begin{figure*}
\includegraphics[width=16 cm]{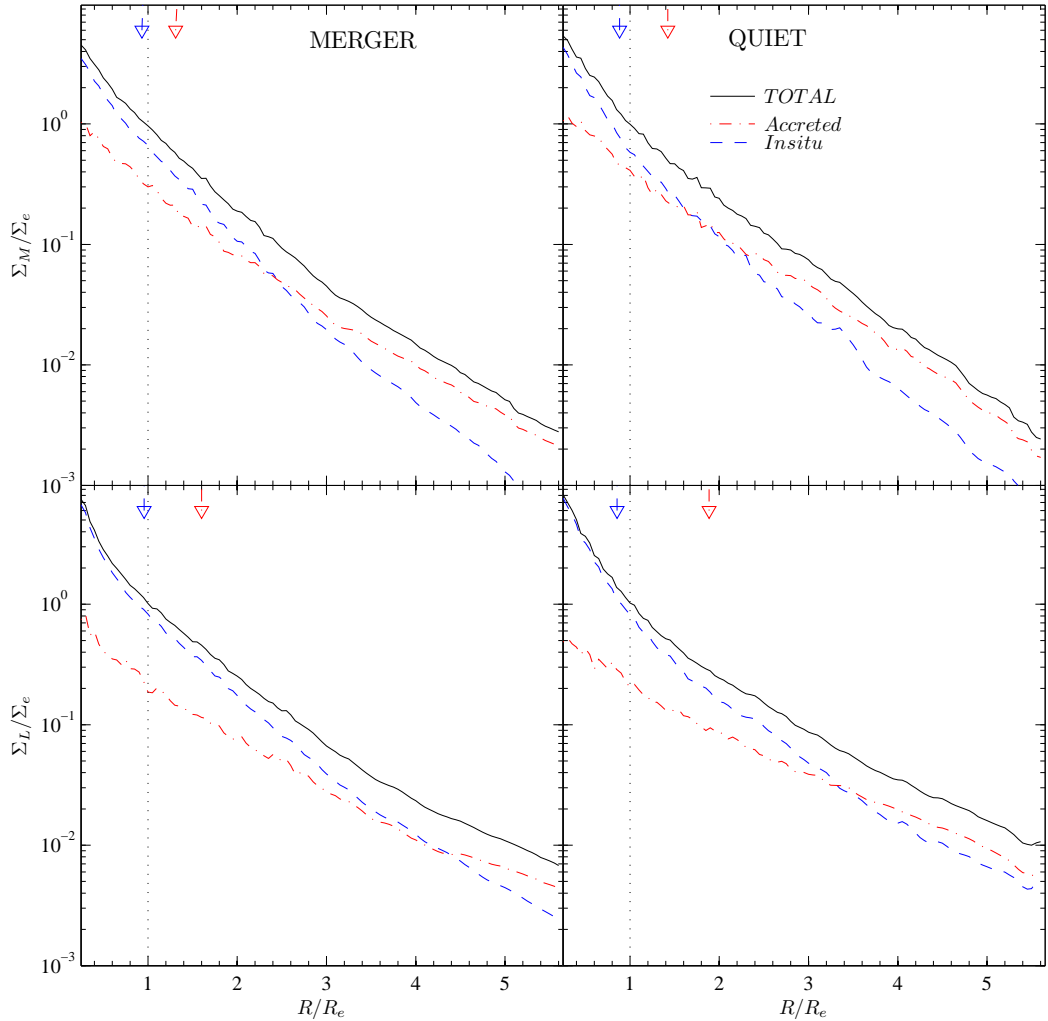}
\vspace{-3cm}
\caption{Mass density (upper panels) and luminosity density (lower
  panels) profiles  for the Merger (left panels) and Quiet (right
  panels) galaxies. The contribution to the total density (black
  solid) is divided
  into the in-situ (blue dashed line ) and accreted (red dot-dashed
  line) components. The arrows indicate the half-mass or effective radii of
  the in-situ and accreted populations in terms of the galaxy
  effective radius.
}
\label{prof_ins}
\end{figure*} 

\begin{figure*}
\includegraphics[width=0.9\columnwidth, angle=-90]{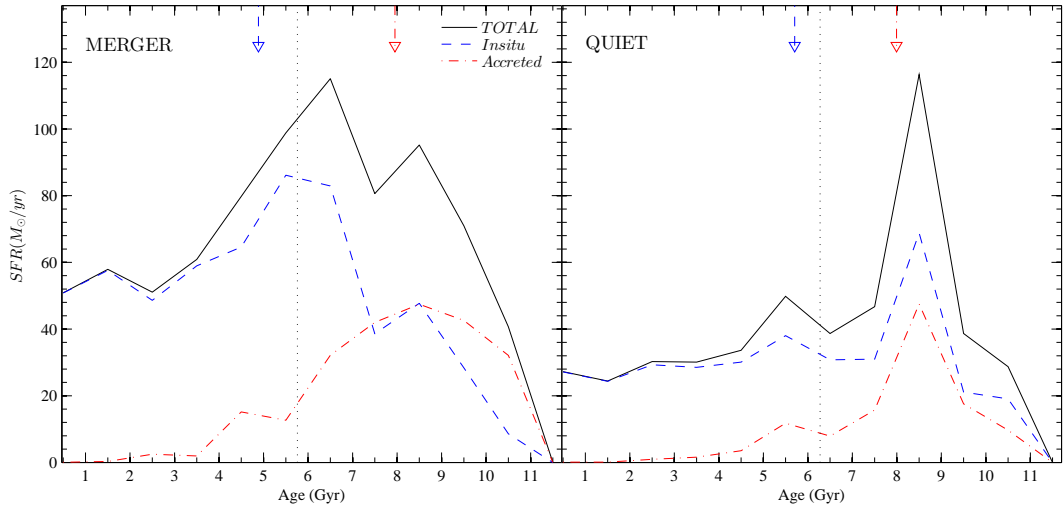}
\caption{Mean star formation histories of galaxies belonging to the M
  (left-hand panel) and  Q samples (right-hand 
  panel). The black solid line indicates the total SFR, the red
  dot-dashed line stays for the accreted component and the blue dashed
  line for the in-situ population.  The dotted vertical line represent
  the mass-weighted mean age of the global population and the arrows indicate the mean ages
  of the in-situ (blue) and accreted (red) components.}
\label{sfh}
\end{figure*} 

\begin{figure*}
\includegraphics[width=1. \columnwidth, angle=-90]{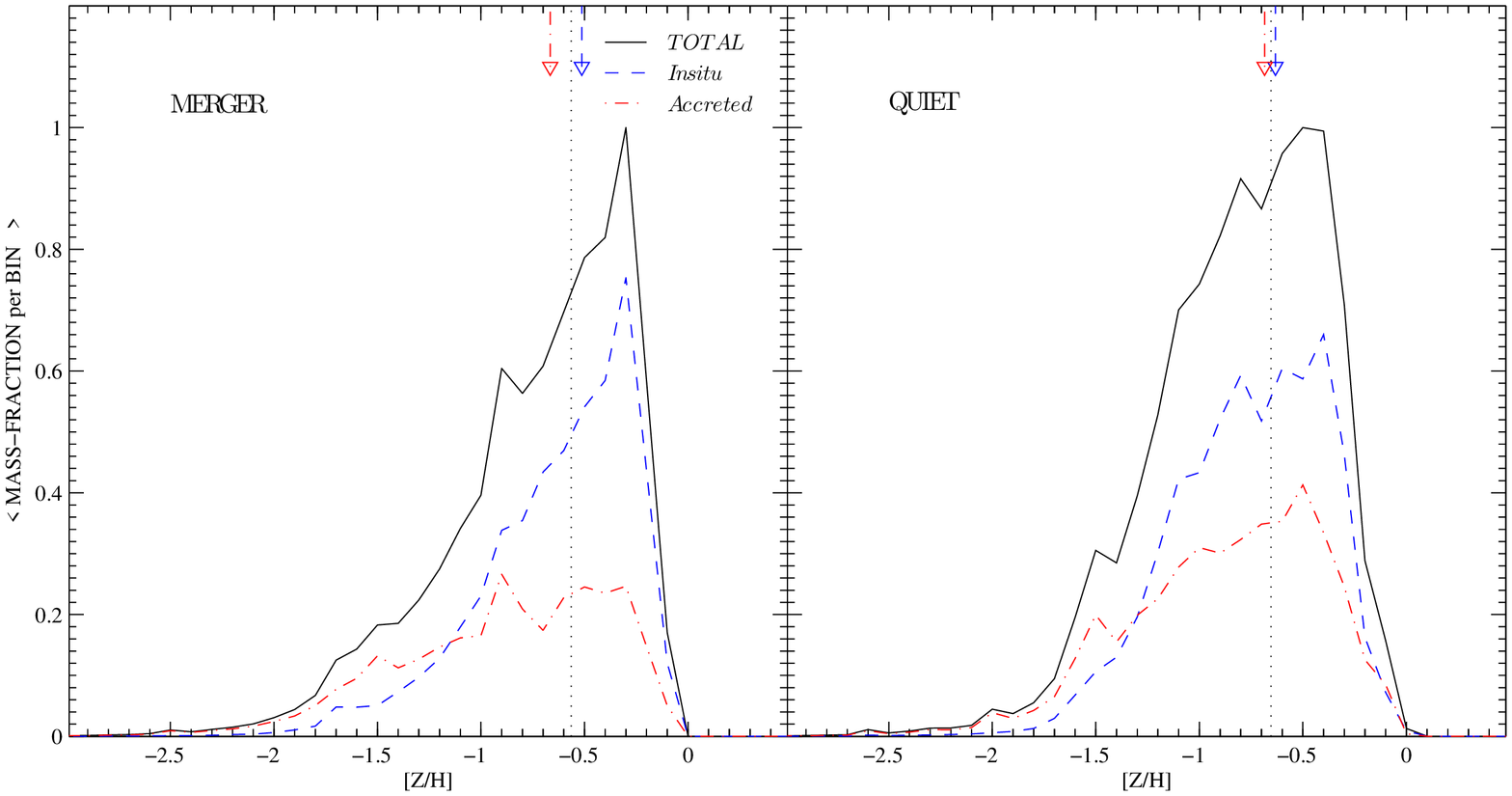}
\caption{Mean metallicity distribution of galaxies belonging to the M
  (left-hand panel) and  Q samples (right-hand 
  panel), normalised to the maximum mass mass-fraction. Lines are as in Figure \ref{sfh}.  The dotted vertical lines and
  arrows indicate the mass-weighted mean metallicity of the global population and of
  the two components.
}
\label{md}
\end{figure*}

According to recent simulations, galaxy evolution seems to be well
described by a two-phase model    \citep{naab09,johansson09,oser10, lackner12}. 
The first phase is a period of in-situ star formation 
and could
resemble the monolithically model of galaxy formation \citep{eggen62,
  larson75}. 
The rest of the stars forming the galaxies are accreted via
mergers, constituting the second phase. In this Section, we study the
differences between these two populations of stars: in-situ and
accreted.  In particular, we focus on their differences concerning the
stellar populations and spatial distribution.

We define as in-situ those stars formed in the main progenitor of
the present-day galaxy, whereas the ex-situ (or accreted) component
are the stars formed outside the main unit and accreted later,
via merger or smooth accretion. 
To differentiate among in-situ and ex-situ stars, we flag
each stellar particle with the halo identifier of the galaxy where it was formed.
Thus, in our definition, all the particles present in the
zero-redshift galaxy and formed in the main progenitor are considered
in-situ, even if they have been lost at some intermediate time-step
and re-accreted later.
To track back the star formation history of the stellar component, we 
follow the merger tree of the
zero-redshift galaxies up to z=4. Since the fraction of stars formed before
this starting redshift is very small, our results on the accretion
history do not depend on this choice. 

In our simulated massive galaxies, the majority of the stars (50-90\%) have been formed in-situ, being the amount of 
accreted stellar mass around 10-50\%. 
Our result roughly agrees with the fraction of accreted mass estimated in the simulation by 
\cite{lackner12}, but contrary to their findings, we do 
 not find any significative dependence of the in-situ fraction with the
final stellar mass of the galaxy. The reason for such difference in the trend of accreted mass 
is likely related with the different approaches used to describe the feedback 
processes and the numerical resolution of the simulation. This is a similar situation to that in \cite{oser10} and \cite{lackner12}, where
both works found quite different trends of the accreted mass with the stellar mass of the galaxies. 
We do find a weak dependence on the merging history, though,
 with the Q galaxies having an higher
fraction of in-situ stars (70\%) than the M galaxies (65\%).
Hence, even galaxies that have not undergone any important merger
event, have a considerable fraction of accreted mass.
These ex-situ stars may have formed in small galaxies ($M<10^{8}-10^{9}
    M_{\odot}$), whose accretion to the main progenitor has not been
    identified as a merger, or formed in very small clumps, not
    recognized as haloes by the halo finder. 

To examine the spatial distribution of the in-situ and ex-situ stars, 
we consider the  M and Q samples. 
The two populations of stars exhibit quite different mass and light
profiles, as shown in Figure \ref{prof_ins}. One of the main features shown by this figure is that  
that the in-situ contribution is clearly dominating the central region of
our simulated objects.
Within the half-mass radius, the in-situ (ex-situ) component represent a fraction 
of 72\% (28\%) of the total mass for the M galaxies and 69\% (31\%) for the Q sub-sample. 
In the case of considering light profiles, the fraction of light within the effective radius from the in-situ (ex-situ) population 
is 84\% (16\%) for the M and 88\% (12\%) for the Q galaxies, respectively. 
The in-situ component dominates in the internal part, but in the
outermost regions the accreted component prevails.
The radius of overtaking lies at $2.5-3 R_e$ in the mass density
profiles and at $3.5-4 R_e$ in the case of luminosity density
profiles. The fact that in luminosity the accreted stars dominate
only in the very external region is due to their old ages that bring
down their luminosities. 
Note that detailed stellar
population studies based on extremely high quality spectra obtained with large
sized telescopes do not reach such galactocentric distance (e.g. \citealt{Sanchez07}). Therefore 
most of the stellar population
studies are in fact sampling the in-situ component, thus leading to a more
uniform view of the stellar content of massive early-type galaxies (e.g. \citealt{renzini06}). 
There is however evidence of a small contribution from a somewhat younger
component on the top of a predominantly old stellar population, mainly when
luminosity-weighting by light. These contributions have been found in various
spectral ranges such as for example in the visible (e.g. \citealt{Trager00}) 
or in the K band \citep{Marmol09}.

The previously discussed fractions are not only 
relevant to constrain better the assembly of
massive early-type galaxies, but could also be linked to their experienced SFHs.
In fact it is becoming increasingly popular to estimate the SFH of these
galaxies employing full spectrum-fitting algorithms (e.g. \citealt{cid05, koleva09}). Although these studies confirm the nearly passively
evolving nature of the bulk of the stellar populations in the central regions of
massive ellipticals, the derived SFHs indicate the presence of different stellar components (e.g. \citealt{rosa11}). 

The half-mass radius of the accreted stars is $1.34 R_{hm}$ for the M galaxies and 
$1.42 R_{hm}$ for the Q galaxies, where $R_{hm}$ is the half-mass
radius of the galaxy. The in-situ population have radii: $0.93 R_{hm}$ and $0.88 R_{hm}$, for M and Q
galaxies, respectively. Thus, 
in both kind of galaxies the accreted stars are found at larger radii
than the in-situ component, with the M galaxies showing a slight 
excess of accreted stars in the central regions compared with the Q
galaxies, that may be ascribed to the stellar mixing
acting during a major merger event.
Considering both samples, M and Q, the half-mass radius of the accreted stars is $\sim 1.5$ larger than the in-situ component.
Such difference in the half-mass radii between the two populations of
stars have been also found by \cite{lackner12}, although the
difference found by these authors is slightly larger than ours.
Similar differences are observed
when looking at the effective radii (bottom panels of Fig.~\ref{prof_ins}).

In Figure~\ref{sfh}, we show the mean star formation history of the galaxies
used in the Q and M samples, separated in the in-situ and accreted
components.  These plots clearly show 
that the accreted stars
are older than the in-situ stars. The ex-situ stars are   $\sim$ 3 Gyr older than 
the in-situ population on average.
In the accreted population the star formation rate ceased about 4  Gyr ago,
whereas the in-situ star formation occurs at significant rate until
the present time.
Since the accreted stars are
typically old,  they are much less luminous than the in-situ stars,
 giving rise to
the strong gap between the luminosity densities of the two populations
shown in the bottom panels of Figure \ref{prof_ins}.
The metallicity distribution for the same galaxies is shown in
Figure~\ref{md}. 
Although the mean metallicity of the two populations is
not very different (the accreted stars are   $\sim 0.1$ dex more
metal-poor than the in-situ population), their metallicity
distributions differ substantially.
The in-situ stars are skewed
toward large metallicities, mainly in the case of the M galaxies, that
show an accumulation of values around $\sim -0.3\, $dex. 
On the contrary, the metallicity distribution of the accreted component
spans a wider range of metallicities, without any characteristic value. This is consistent with the fact that the ex-situ stars formed in
galaxies having a variety of mass and, hence, metallicity. The results in Fig.\ref{md} show averaged metallicities over the whole galaxy, and therefore, they cannot be directly compared with the  
Fig. \ref{history} where median metallicities radial profiles are presented. The inclusion of the external parts of the galaxies in the calculation of the mean metallicities produces slightly lower values of these quantities.

The analysis of the Fig.~\ref{sfh}  revels that the star formation in our simulation is suppressed at epochs before z$\sim$4.
The reason for that is purely numerical and it has to do with the fact that 
the AMR approach used in the present work refines the grid when the mass (gas or dark matter) in a cell is above a certain threshold (see Sec. 2.1). Given that we  consider a cosmological box of 44 Mpc and the maximum number of levels is seven, the numerical resolution is moderate ($\sim$ 2.7 Mpc).  In practice, this implies that the most nonlinear structures are smoothed and their evolutions are not properly described. As the creation of highest levels of refinement is directly linked with the nonlinear growth of the structures, these 
high resolution patches appear at later times. Let us point out that in our approach the star formation can only take place at the highest levels of refinement, and therefore, the star formation in our simulation is delayed due to the numerical effects associated to the late creation of high resolution patches. This would be the reason why the star formation begins at z$\sim$4.

\section{Summary and discussion}\label{discussion}

In this paper we present the results of a cosmological AMR simulation including cooling and heating processes as well as 
a phenomenological star formation  and type II supernovae feedback. The computational domain is a cosmological box of $44\, Mpc$ side
length. In this volume, and by means of an adaptive friends-of-friends
algorithm applied to the stars, we identify a sample of virtual
galaxies.

Our simulation  does not consider any resimulation  and therefore, the whole computational box is treated 
consistently. An important drawback of this approach is that, despite the use of an AMR technique, when starting from a cosmological box, the numerical resolution is still limited. Besides, due to the intrinsic nature of our AMR algorithm, designed to refine high density regions, our simulation is biased to better describe the massive objects. Keeping in mind these constraints, we use the galaxy catalogue extracted from the simulation to study the properties of the massive galaxies nowadays, and the processes responsible of their actual properties.

We find  33 galaxies in the simulation with stellar masses larger that 
$M_* > 10^{11}\,M_{\odot}$. This tentative sample of galaxies is filtered with some extra conditions. Thus, 
we only consider galaxies that are located at the highest level of refinement, therefore being resolved with the best numerical 
resolution, and galaxies that have not undergone any recent merger event or star formation outburst. These 
restrictions lead us to a final sample including 21 massive galaxies.
We are aware that our simulated galaxies are overly massive
  compared with the observed ones. Indeed, the baryonic conversion
  efficiency is $\sim 4$  times higher than that expected from
  abundance matching techniques.  We ascribe the main reason for such
  overproduction of stars to the lack of
  AGN feedback, that is expected to quench the late star formation in
  massive galaxies. For this reason we have considered post-processing
  corrections to the stellar mass in order to bring down the
  discrepancy with observations.

In order to analyze our sample, we classify the galaxies according to
three criteria: i) dynamical properties, ii) evolutionary history and,
iii) morphologies. Adopting the first criterion, galaxies are
classified as slow rotators (54\%) or fast rotators (46\%) according to the prescription by \cite{emsellem11}. In the second case, the galaxies are separated in those ones that have undergone at least a merger event (52\%) and those 
ones that have had a quite evolution (48\%). Finally, the sample is divided into disk-like (30\%) and elliptical (70\%) galaxies according to their 
S\'ersic index. For any of the sub-samples produced with the previously mentioned criteria, we produce 1D profiles of the following 
quantities: luminosity $\Sigma_L$, surface density $\Sigma_M$ , velocity dispersion $\sigma $, line of sight velocity $v$, age and metallicity $Z$.

Consistently with observations we find that most of the massive galaxies
have an early-type morphology. The median profile of the elliptical
galaxies is well fitted by a S\'ersic model having index $n=3.5$.
However, this value is 
somehow lower than that of typical ellipticals in the nearby universe,
which show $n \geq 4$.
Together with the fact that we cannot reproduce the old nature of
elliptical galaxies,  it might indicate that the resolution of our simulation
is still marginal to correctly model the morphology of our
galaxies. 

Concerning the dynamics, we find that 
the velocity dispersion profiles are almost flat within the effective
radius and decay in the outer region. In the case of disc galaxies
and merging galaxies,  the velocity
dispersion falls rapidly outside the effective radius, while in the
other case the decaying is milder.  
The rotation curves show also a  dependence on the
merging history of the galaxies, thus, the galaxies that have undergone mergers
show high rotational velocities out to the external regions.

All the subsamples studied  
 present positive age gradients, significant
mainly within the effective radius,  and negative
metallicity gradients. 
As for the dynamics, the most important factor in establishing the stellar population
gradients turns out to be the merging history. Galaxies having 
undergone mergers are younger and more metal-rich than the quiet
galaxies and present steeper metallicity profiles.
Since most of the mergers occurring in the simulation entail a
significant enhance in the star formation, the rejuvenation of the
stellar population is responsible for such young  and metal-rich
stars, mainly in the central part of the galaxies.

We also study the star formation history in the galaxies of the sample. The stellar build-up of the galaxies is established by two mechanisms. The first one is the formation of stars within the galaxy, or its main progenitor backwards in time. These are the stars formed in-situ. 
The second mechanism increases the stellar mass of the galaxy through
mergers 
 or by smooth accretion of stars in the galactic halo. We called this population the ex-situ or accreted stars. 
The majority of the stars formed in the simulated galaxies are formed in-situ, representing between the 50\% and the 90\% of the 
stellar mass of the galaxies. The in-situ star formation process is more intense at early epochs of the life of the galaxies. This active 
phase lasts typically around a couple of Gyrs, but the process does not stop suddenly, and it keeps on forming stars within the galaxy 
-- at a minor rate though -- until present time. 
The ex-situ star formation, on the contrary, takes place mainly at
early times and becomes very marginal at
low redshift.

Given the leading role of mergers in determining the most significant
features of the galaxies in the simulation, we explore the dependence
of the in-situ and ex-situ stars on the merging history. As expected,
the merger galaxies have a higher fraction of accreted stars,
though the in-situ stars are always the dominant population in the inner parts.
In both, the merger and the quiet galaxies, we find an important difference in the spatial
distribution of the in-situ and ex-situ stars, leading to quite
different light and mass density profiles. The in-situ stars dominate up to a
few effective radii, 
whereas there is an 
overtaking of the accreted component in the outermost regions.
The effective radii (or half-mass in the case of mass density
profiles) of the two components are clearly different.
The effective radii of the accreted stars are always larger (by $1.5$
times) than those
of the in-situ component, highlighting the fact that the ex-situ stars are mainly located in outer regions compared with the in-situ stars.
Although the trend for the merger and quiet galaxies is similar, the
merger galaxies present a slightly excess of accreted stars in the central region compared with the quiet galaxies. This 
situation would be caused by the mixing action of the merger events.

The analysis of the star formation histories of both, M and Q
galaxies, shows that accreted stars are always older ( $\sim 3$ 
 Gyr) and on average less metallic ($0.1$ dex) than the in-situ stars. This
explains the changes in the luminosity density and mass density for
the same galaxies, as the two components look like differently
depending whether their mass or  luminosity are considered. The accreted
population also shows a much larger dispersion in the metallicity
distribution, indicating that they formed in a variety of systems.

The numerical resolution of the simulations is a crucial issue when understanding some of the results we find. The fact that a unique simulation is performed without resimulations produces that some of the regions are not optimally described, specially small haloes. On the other hand, the lack of resolution can be interpreted as an uncontrolled source of numerical feedback. This is the reason why the star formation rates -- at present time -- in our galaxies are slightly higher than expected. The lower resolution delays or sometimes prevents the star formation at early epochs leaving more gas available for a more extended star formation history. 

Related with the use of a lower resolution in some regions, low mass
galaxies are less favored and therefore, the number of such objects is
low. Consequently, the number of minor mergers may be lower than
expected. 
The low merger rate together with the excess of the star formation at
late times imply  that the estimate of the accreted fraction in our
simulation should be considered as a lower limit.
We plan to explore these issues by means of higher resolution
simulations in the next future.

\section*{Acknowledgements} 
We thank the referee for enlightening  comments and constructive
criticism. We are also grateful to  I. Trujillo and S. Planelles for useful discussions. This work was carried out within the Coordinate Project RAVET and it was  supported by the Spanish Ministerio de Econom\'{\i}a y Competitividad 
(MINECO, grants   AYA2010-21322-C03-01,
AYA2010-21322-C03-02)   and    the   Generalitat   Valenciana   (grant
PROMETEO-2009-103).   JNG thanks  to  the MICINN  for  a FPI  doctoral
fellowship.   Simulations   were  carried  out  at   the  {\it  Servei
d'Inform\'atica de la Universitat de Val\`encia}.

\appendix
\section{Velocity dispersion estimates}

In order to assess the reliability of the velocity dispersion
measurements in our simulated galaxies, we have compared the dynamical 
method to measure the velocity dispersions with an approach similar to
that performed in the observations
(e.g. \citet{2006MNRAS.366.1126C,emsellem11}). 
In such an approach, to derive a 
spectroscopic velocity dispersion, 
 we use the broadening of
the spectral features.

For each cell of the artificial images a spectrum has been built
by integrating over all the particles in the cell, shifted in wavelength
by their velocity relative to the center of mass of the
galaxy:
\begin{equation}
F_{cell}(\lambda)=\sum_{i=1}^{Ncell}(m_i F(t_i, Z_i, \lambda(1+v_i/c)) (1+v_i/c))
\end{equation}
where $F_{cell}$ is the integrated flux in the cell, Ncell is the number of particles in the cell, $m_i$ is the
stellar mass of the i-th particle, $t_i$ and $Z_i$ are their age and
metallicity, $v_i$ is its radial velocity relative to the center of
mass, and c the speed of light. 
$F(t_i, Z_i, \lambda(1+v_i/c))$ is the flux of the single stellar
population model having the same age and metallicity of the stellar particle and shifted in wavelength
according to the particle velocity. 
For the purpose of producing the broadened spectra, we have used the
MIUSCAT stellar population models in the range 6400-7400\,\AA,
where the mean  resolution of the models is 46 km/s and the velocity
scale is 39 km/s. The use of a more extended 
wavelength range of the MIUSCAT models (3500 - 7400\, \AA) does
not significantly change the results.

The line of sight velocity distribution (LOSVD) has been recovered from the
comparison with  a model spectrum created by convolving a template with a
parameterized LOSVD.
To minimize the template mismatches, we have used as a template
spectrum the integrated spectral energy distribution in a given cell without applying the
velocity shifts, hence at the nominal resolution of the models.
As the fitting algorithm, we have used the Penalized Pixel Fitting method
(PPxF; \citealt{2004PASP..116..138C}) fitting the first two moments: V
and $\sigma$.  The higher-order Gauss-Hermite moments are set to zero
and are not considered in the fit.
To have an acceptable sampling of the velocity distribution, the
procedure has been performed only in those cells with at least 1000
particles. The final estimates of $\sigma_{spec}$ is given by the velocity dispersion measured
from the spectrum corrected by the model resolution. 

\begin{figure*} 
\includegraphics[angle=-90,width=1.9\columnwidth]{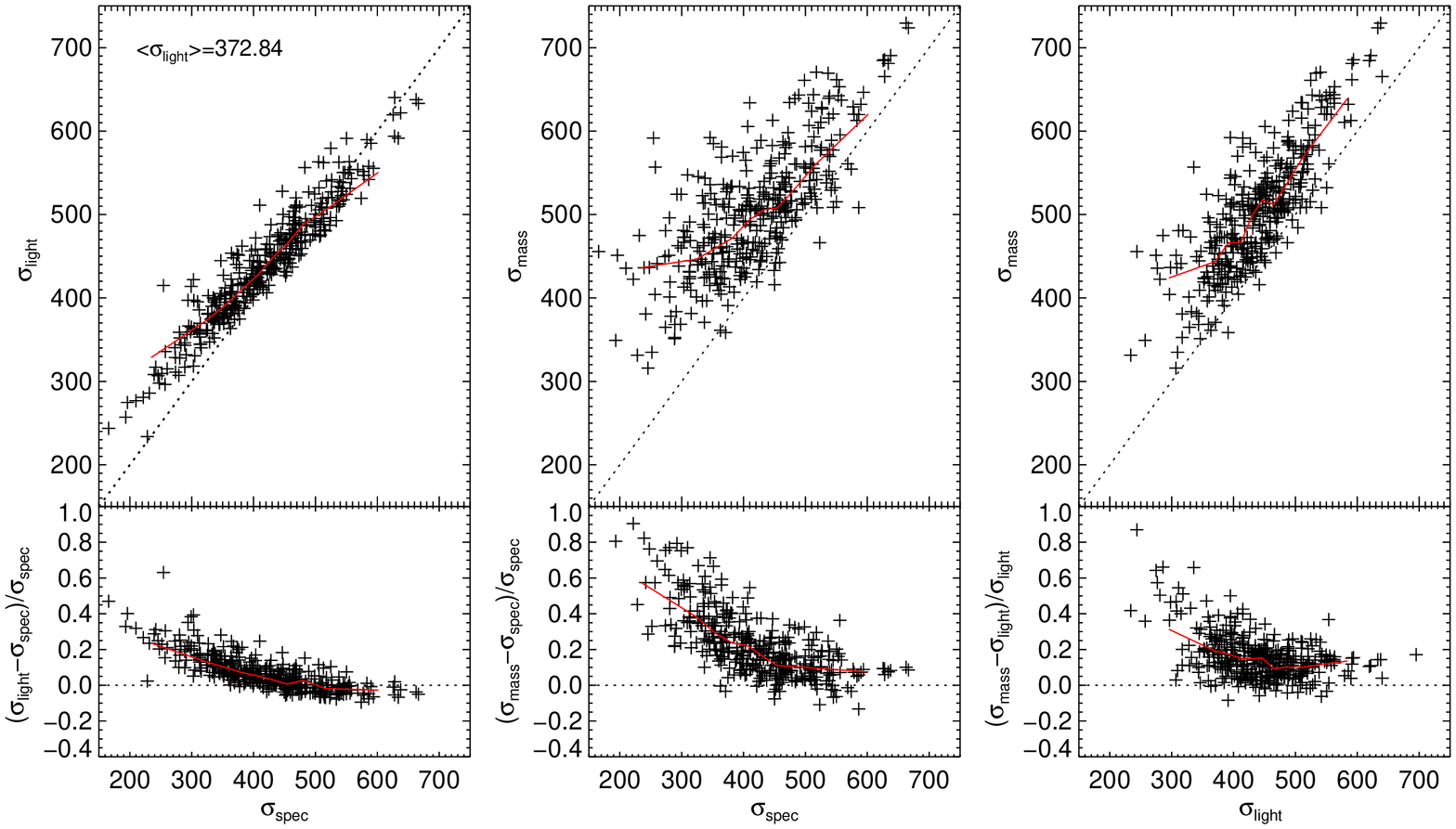}
\caption{Comparison between dynamical and spectroscopic determinations 
  of the velocity dispersion for all the cells with at least 1000
  particles of one massive galaxy. The left-hand panels show the
    comparison between the dynamical velocity dispersion weighted by
    light, $\sigma_{light}$, and the spectroscopic velocity dispersion
    $\sigma_{spec}$. The middle panels show the comparison between the
    dynamical determination of the velocity dispersion weighted by mass , $\sigma_{mass}$, and the
    spectroscopic one, whereas the right-hand panels show the
    comparison between the two dynamical estimates, $\sigma_{mass}$
    and $\sigma_{light}$. All the  three
  projections of the galaxy are shown.  The upper panels show the
  velocity dispersion comparisons
  whereas the bottom ones display the
  residuals. Each cross is a single determination in one cell. The
  dotted lines indicate the one to one relation and the red solid
  lines 
  show the median values computed in equally populated intervals of velocity
  dispersion. 
}
\label{sigma1} 
\end{figure*}

In Figure \ref{sigma1} we show the comparison between the velocity
dispersion derived from the broadened spectra and and that from two different
 dynamical estimates, for one of the massive galaxies of our sample. 
A mass-weighted velocity dispersion, $\sigma_{mass}$, is defined as 
the square root of
 the mass-weighted mean of the squared deviations of the particle
 velocities:
\begin{equation}
\sigma_{mass}^2= \frac{\sum\limits_{i=1}^{Ncell}m_i(v_i-v_{cell,mw})^2 } {\sum\limits_{i=1}^{Ncell}m_i} 
\end{equation}\label{sigma_mw}
where $v_{cell,mw}$ is the mass-weighted mean of the particle
velocities in the cell. An analogous way of defining the dynamical
velocity dispersion is by weighting in  luminosity instead of mass:
\begin{equation}\label{sigma_lm}
\sigma_{light}^2=\frac{\sum\limits_{i=1}^{Ncell}L_i(v_i-v_{cell,lw})^2 } {\sum\limits_{i=1}^{Ncell}L_i} 
\end{equation}
where $L_i$ is the particle luminosity in the r-band and $v_{cell,lw}$
is the luminosity-weighted mean of the velocities in the cell.

At low velocity dispersions the spectroscopic determinations 
significantly deviate from both the dynamical values, being the
discrepancy stronger with the mass-weighted velocity dispersion.
It is worth to note that when the velocity dispersion is low, the spectroscopic measurements are less
reliable, because approaching to the resolution limit of the
models. Indeed, galaxies having the lowest mean velocity dispersions show
the strongest deviations.  Another important source of uncertainty in the
spectroscopic determination is the poor sampling of the velocity
distributions. In several cases, the velocity distribution is far from
being  gaussian and, mainly for galaxies with significant rotation, the projection
effects can dramatically alter its shape,
giving rise to strongly asymmetric distributions, that cannot
be well parametrized by the LOSVD.  
Such an effect can become dramatic in the outer region of the galaxy, but 
it is not affecting too much the cells within the effective radius.

When the velocity dispersion is high, the spectroscopic value correlates well with the dynamical
one, although the correlation is much tighter for
$\sigma_{light}$ than $\sigma_{mass}$. 
As expected, the spectroscopic velocity dispersion better correlates
with the luminosity-weighted velocity dispersion. Indeed, in the 
integration of the SSP spectra, the velocities of the most luminous
particles give the largest contribution to the broadening of the
spectra. The mass-weighted velocity dispersion displays a higher
scatter and a bias towards higher values. The same trend is visible in
the comparison between $\sigma_{mass}$ and $\sigma_{light}$ (third panel of Figure \ref{sigma1}).
The reason for the discrepancy between the two dynamical estimates can be understood by looking at the velocity
dispersion of old and young stellar populations,
with the oldest population  ($>$5  Gyr) being the one with the higher
dispersion in velocity.
Since the oldest stellar particles  in the simulation have been formed at the
beginning of their life through strong starbursts, leading to high mass
particles, their velocity deviations  weigh more on the computation of
$\sigma_{mass}$, bringing it towards high values.
On the other hand, when the velocity deviations are weighted by the
particle luminosity, the weight of the massive particles is smoothed
down because of their high M/L.

\begin{figure*} 
\includegraphics[angle=-90,width=1.9\columnwidth]{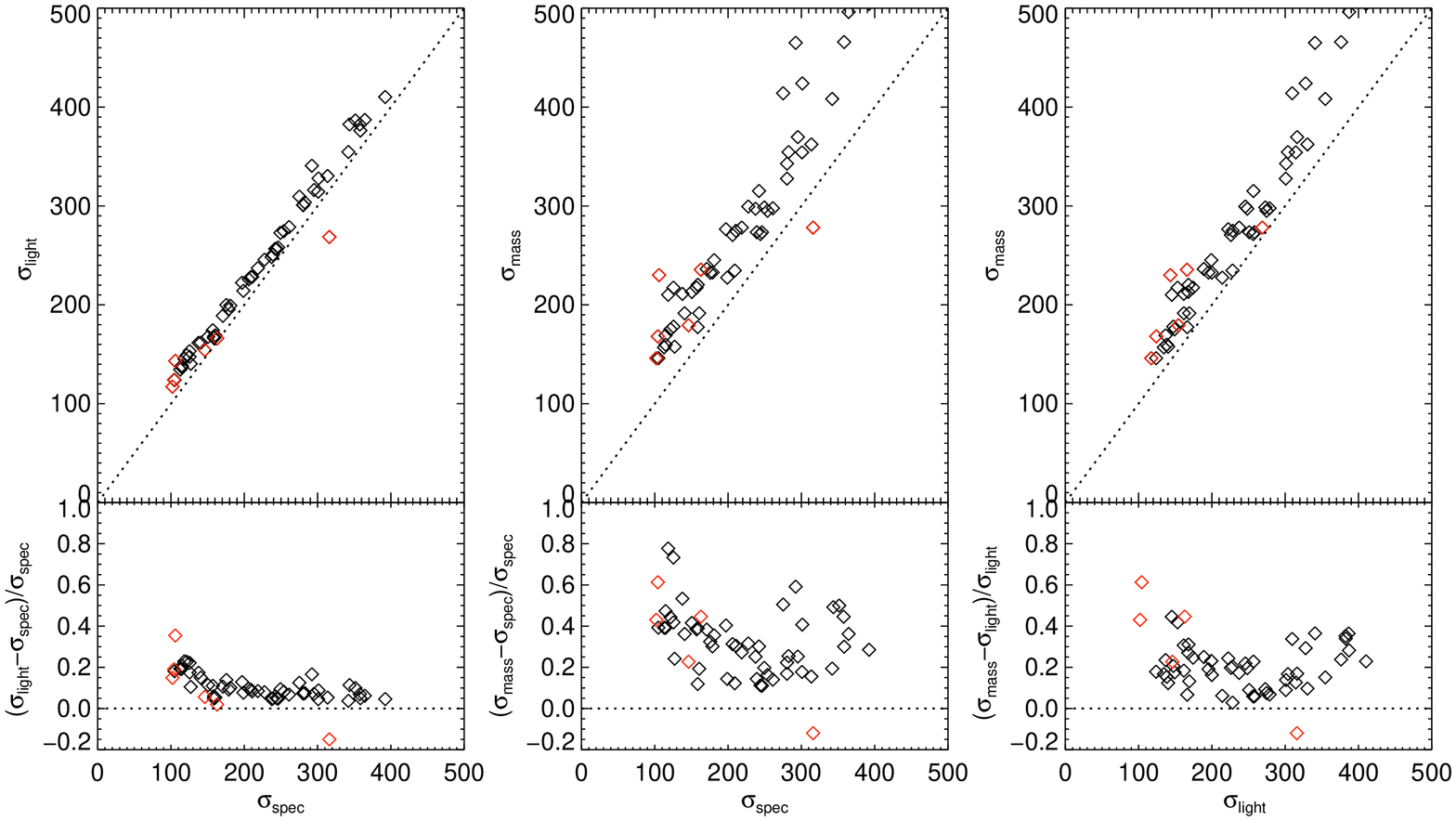}
\caption{Comparison between dynamical and spectroscopic average 
  velocity dispersions within one effective radius 
  of the massive galaxies in our sample at z=0. Galaxies in all the
  three projections are shown.
  The upper panels show the comparisons
  between the velocity dispersions whereas the bottom ones display the
  residuals. Red symbols indicate galaxies for which less than 10 cells
  have been used for computing the average. 
 The  dotted lines indicate the one to one relation. 
}
\label{sigma_all} 
\end{figure*}

The effect on the integrated velocity dispersion of each galaxy is
shown in Figure \ref{sigma_all}. The galaxy velocity dispersion has
been computed by averaging all the velocity dispersion values in the
cells within the galaxy effective radius.
The cells that do not have a sufficient  number of
particles or whose spectroscopic velocity dispersion falls below the
model resolution have been not taken into account in the average.
Galaxies having less than ten  cells satisfying this condition have been denoted with
red symbols in Figure \ref{sigma_all}. 
The luminosity-weighted velocity dispersion shows a tight correlation
with the spectroscopic determination, although  they tend to be
higher by $\simeq$10\%. This systematic shift is driven by the contribution of low-velocity
dispersion cells whose dynamical velocity dispersions are biased towards high values.
On the other hand, the mass-weighted velocity dispersion still shows a
larger scatter and a strong bias towards high values, even in the high
velocity dispersion range.

Since in the spectroscopic determination of the velocity
dispersion we have followed the same approach used in the
observations, we rely on the dynamical determination closer to
$\sigma_{spec}$, hence in our analysis we have used $\sigma_{light}$
instead of $\sigma_{mass}$.  

\end{document}